\documentclass[aps,prb,10pt,twocolumn,superscriptaddress,longbibliography]{revtex4}
\usepackage{graphicx}
\usepackage{amssymb, amsmath}
\usepackage{color}
\usepackage{ulem}
\usepackage{bm}
\usepackage{bbm}
\usepackage{graphicx}
\usepackage{subfigure}
\usepackage{dcolumn}
\usepackage{mathtools}
\usepackage{pifont}
\usepackage{diagbox}
\usepackage{braket}
\usepackage{verbatim}
\usepackage{diagbox}

\def\I{\uppercase\expandafter{\romannumeral 1}}
\def\II{\uppercase\expandafter{\romannumeral 2}}
\def\III{{\uppercase\expandafter{\romannumeral 3}}}
\def\IV{{\uppercase\expandafter{\romannumeral 4}}}
\def\V{{\uppercase\expandafter{\romannumeral 5}}}
\def\VI{{\uppercase\expandafter{\romannumeral 6}}}
\def\VII{{\uppercase\expandafter{\romannumeral 7}}}
\def\i{\lowercase\expandafter{\romannumeral 1}}
\def\ii{\lowercase\expandafter{\romannumeral 2}}
\def\iii{{\lowercase\expandafter{\romannumeral 3}}}
\def\iv{{\lowercase\expandafter{\romannumeral 4}}}
\def\v{{\lowercase\expandafter{\romannumeral 5}}}
\def\vi{{\lowercase\expandafter{\romannumeral 6}}}
\def\vii{{\lowercase\expandafter{\romannumeral 7}}}

\def\angstrom{\mbox{\normalfont\AA}}

\def\nn{\nonumber\\}

\def\angstrom{\mbox{\normalfont\AA}}

\def\nn{\nonumber\\}

\def\s{\textrm{s}}

\begin{document}

\title{Alternating twisted mutilayer graphene:  generic partition rules, double flat bands, and orbital magnetoelectric effect}% Force line breaks with \\

\author{Bo Xie}
\affiliation{School of Physical Science and Technology, Shanghaitech University, Shanghai 200031, China}%

\author{Shihao Zhang}
\affiliation{School of Physical Science and Technology, Shanghaitech University, Shanghai 200031, China}%

\author{Jianpeng Liu}
\affiliation{School of Physical Science and Technology, Shanghaitech University, Shanghai 200031, China}%
\affiliation{ShanghaiTech laboratory for topological physics, ShanghaiTech University, Shanghai 200031, China}

\date{\today}% It is always \today, today,

\begin{abstract}
Twisted graphene systems have draw significant attention due to the discoveries of various correlated and topological phases. In particular, recently the alternating twisted trilayer graphene is discovered to exhibit unconventional superconductivity, which motivates us to study the electronic structures and  possible interesting correlation effects of this class of alternating twisted graphene systems. In this work we consider generic alternating twisted multilayer graphene (ATMG) systems with $M$-$L$-$N$ stacking configurations, in which the $M$ ($L$) graphene layers and the $L$ ($N$) layers are twisted by an angle $\theta$ (-$\theta$). Based on   analysis from a simplified $\textbf{k}\!\cdot\!\textbf{p}$ model approach, we analytically derive generic partition rules for the low-energy electronic structures, which exhibit various intriguing band dispersions including one pair of flat bands, two pairs of flat bands, as well as flat bands co-existing with with Dirac cones, quadratic bands, or more generally  $E(\mathbf{k})\!\sim\!k^J$ dispersions ($J$ is positive integer) for each spin and valley. Such unusual non-interacting electronic structures may have unconventional correlation effects.  Especially for a mirror symmetric ATMG system with two pairs of flat bands (per spin per valley), we find that Coulomb interactions may drive the system into a state breaking both time-reversal and mirror symmetries, which can  exhibit a novel type of orbital magnetoelectric effect due to the interwining of electric polarization and orbital magnetization orders in the symmetry-breaking state.
\end{abstract}

\maketitle

%%%%%%%%%%%%%%%%%%%%%%%%%%%%%%%%%%%%%%%%%%%%%%%%%%%%%%%%%%%%%%%%%
%%%%%%%%%%%%%   introduction  %%%%%%%%%%%%%%%%%%%%%%%%%%%%%%%%%%%
%%%%%%%%%%%%%%%%%%%%%%%%%%%%%%%%%%%%%%%%%%%%%%%%%%%%%%%%%%%%%%%%%

The recent discoveries of some intriguing phenomena, such as  superconductivity\cite{cao-nature18-supercond,dean-tbg-science19,marc-tbg-19, efetov-nature19,efetov-nature20,young-tbg-np20,li-tbg-science21,cao-tbg-nematic-science21},  quantum anomalous Hall effect\cite{young-tbg-science19, sharpe-science-19, efetov-arxiv20,yazdani-tbg-chern-arxiv20,andrei-tbg-chern-nm21,efetov-tbg-chern-arxiv20,pablo-tbg-chern-arxiv21}, and correlated insulator states\cite{zhang-tdbg-np20,cao-nature18-mott,efetov-nature19,tbg-stm-pasupathy19,tbg-stm-andrei19,tbg-stm-yazdani19, tbg-stm-caltech19, young-tbg-science19,efetov-nature20,young-tbg-np20,li-tbg-science21} in magic-angle twisted bilayer graphene(TBG) has aroused great interest. In magic-angle TBG \cite{BM}, the interlayer moir\'e potential generates  pseudo magnetic fields, which are coupled with the Driac fermions from the two layers, leading to topological non-trivial flat bands with eight-fold degeneracy with valley, spin and sublattice degrees of freedom \cite{ashvin-analytical-tbg,liu-pll,song-tbg-prl19, yang-tbg-prx19,po-tbg-prb19,ashvin-prr20}. Such degeneracy can be split by strong Coulomb interactions, leading to symmetry-breaking states reminiscent of quantum Hall ferromagnetism. The interplay between non-trivial topology and strong  Coulomb interaction give rise to fruitful physics in magic angle TBG \cite{balents-review-tbg,andrei-review-tbg,jpliu-nrp21,kang-tbg-prl19,Uchoa-ferroMott-prl,xie-tbg-2018, wu-chiral-tbg-prb19,  zaletel-tbg-2019, wu-tbg-collective-prl20, zaletel-tbg-hf-prx20,jpliu-tbghf-prb21,zhang-tbghf-arxiv20,hejazi-tbg-hf,kang-tbg-dmrg-prb20,kang-tbg-topomott,yang-tbg-arxiv20,meng-tbg-arxiv20,Bernevig-tbg3-arxiv20,Lian-tbg4-arxiv20,regnault-tbg-ed,zaletel-dmrg-prb20,macdonald-tbg-ed-arxiv21,meng-tbg-qmc-cpl21,lee-tbg-qmc-arxiv21,bultinck-tbg-strain-prl21,vic-nc20,zhu-prl20,huang-prl21,balents-prb21}.

The intriguing flat-bands physics is not unique for magic-angle TBG. It has been theoretically proposed and experimentally observed that topologically nontrivial flat bands with strong correlation effects can also  exist in twisted multilayer systems  such as twisted bilayer-monolayer graphene and twisted double bilayer graphene \cite{jpliu-tmg-prx,Ashvin-nc19,koshino-tdbg-prb,kim-tdbg-nature20, cao-tdbg-nature20, zhang-tdbg-np20,young-monobi-nature20,Yankowitz-monobi-np2020,Yankowitz-doublebi-np2020,shi-tbmg-np21,ma2021doubled}. Moreover, recently unconventional superconductivity has been observed in alternating twist trilayer graphene \cite{kim2021spectroscopic,cao2021pauli,park2021tunable,liu2021coulomb}, which is a new type of twisted graphene system with topologically nontrivial flat bands co-existing with dispersive Dirac cone  around the charge neutrality point (CNP) \cite{eslam-prb19,li-arxiv19}. This motivates us study the electronic structures and correlation effects of alternating twisted multilayer graphene systems. One would expect that the extra twist may fundamentally change the low-energy electronic structures, and the extra layers  may introduce additional degrees of freedom that may give rise to versatile interaction effects \cite{scheurer-arxiv21,lian-trilayer2-prb21,bernevig-trilayer1-prb21}
% both experimentally and theoretically proposed that correlated states and unconventional superconductivity can also exist in alternating twisted trilayer graphene\cite{kim2021spectroscopic,cao2021pauli,park2021tunable,liu2021coulomb,zuo2018scanning, li-arxiv19,lei-prb21,scheurer-arxiv21,bernevig-trilayer-prb21,bernevig2021ttghf,bernevig2021ttg3,amorim2018electronic,fischer2021unconventional,wei2021inplane} as well as non-trivial topology, superconductivity, correlated insulator and other symmetry breaking states in twisted multilayer graphene\cite{jpliu-tmg-prx,mora2019flatbands,lu2021valley,khalaf2019magic,he2021chirality,kammermeier2019plane,cao2021ab,tong2021spectroscopic,xu2021tunable}. There might be more dgrees of freedom leading to different states  after introducing more layers. This motivates us study the electronic structures, topological properties, and correlation effects of alternating twisted multilayer graphene systems. Espceially for those have mirror($m_z$) symmetry, the single particle and interaction properties may be influenced by $m_z$ symmetry.

In this work, we theoretically study alternating twisted multilayer graphene(ATMG) : a class of twisted graphene consisting of three sequences of stacking graphene with alternating twist angle, denoted as $M-L-N$. We describe the low energy physics in the non-interaction regime with a continuum model. We classify them by low energy band dispersion for each spin and valley into three types, including one pair of flat bands, one pair of flat bands co-exist with Dirac cone(quadratic bands) and $E(\mathbf{k})\sim k^{J}$($J$ is positive integer), and two pairs of flat bands co-exist with $E(\mathbf{k})\sim k^{J}$($J$ is positive integer). Based on an analytic analysis from a simplified $\textbf{k}\!\cdot\!\textbf{p}$ model approach, which exclude the redundancy parts of the continuum model, we find that the low energy band structure can be described by a generic partition rule. % The generic partition rules indicate that there are three kinds of band dispresions for one valley and one spin. %:  one pair of flat bands with large band width, one pair of flat bands co-exist with Dirac cone(quadratic bands) and $E(\mathbf{k})\sim k^{J}$($J$ is positive integer), and two pairs of flat bands co-exist with $E(\mathbf{k})\sim k^{J}$($J$ is positive integer). Then we construct a simplified $\textbf{k}\cdot \textbf{p}$ modle,which exclude the redundancy parts of the continuum model . The simplified $\textbf{k}\cdot \textbf{p}$ modle describes the low energy band dispersion well and can explain the partition rules. 
According to the partition rule, there must be double flat bands for ATMG  with mirror $m_z$ symmetry and more than one layers in the middle sequence. %We take $A-ABA-A$('$ABA$' is Bernal stacking) as a special case. 
Finally, we consider Coulomb interaction effects and screening effects from remote bands in $A-ABA-A$(with two pairs of flat bands). We study the ground states at different integer filling factors under zero external field. It turns out that the mirror symmetry can be broken spontaneously by Coulomb interaction. % With a Hartree-Fock apporximation and constrained Random Phase Apporximation, we display the ground state under different integer filling factor. 
%We also perform calculations under finite vertical electric field to study the polarization and orbital magnetization response.
%We evaluate the response under vertical electric field and magnetic field through polarization and orbital magnetization. 
Besides, a calculation under finite vertical electric field indicates that% the response is dominated by vertical magnetic field, and 
the electric field can enhance the orbital magnetization linearly.

%%%%%%%%%%%%%%%%%%%%%%%%%%%%%%%%%%%%%%%%%%%%%%%%%%%%%%%%%%%
%%%%%%%%%%%%   set up  %%%%%%%%%%%%%%%%%%%%%%%%%%%%%%%%%%%%%%
%%%%%%%%%%%%%%%%%%%%%%%%%%%%%%%%%%%%%%%%%%%%%%%%%%%%%%%%%%%
%fig1
\begin{figure}[b]
\begin{center}
    \includegraphics[width=8.5cm]{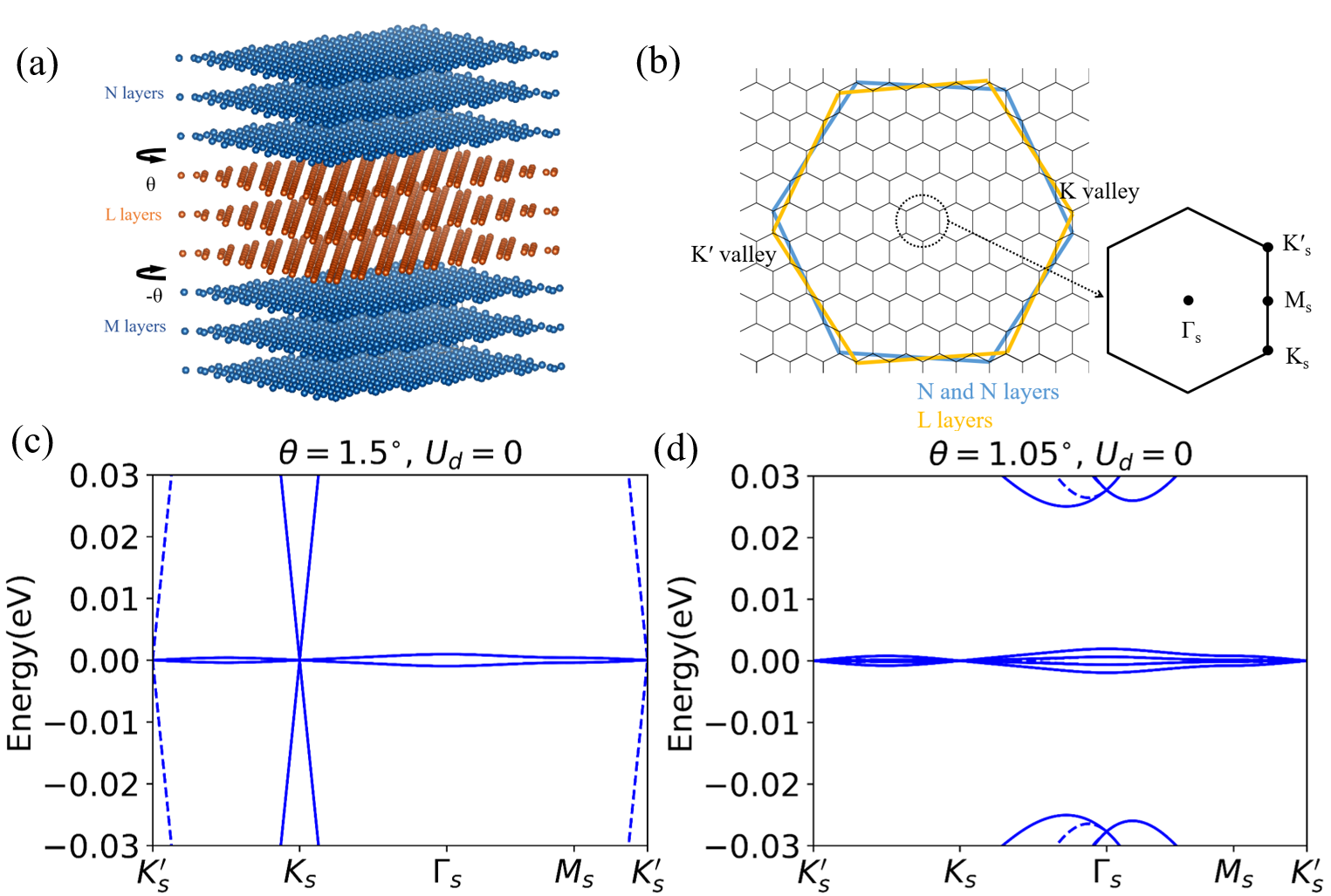}
\caption{
(a) Lattice structure of alternating twist multilayer graphene. (b) Brillouin zones of $M$, $L$ and $N$ layers and Moir\'e Brillouin zone with high symmetry points. Band structure of (c) $A-A-A$ and (d) $A-ABA-A$ at $K$ valley(solid line) and $K^{\prime}$ valley(dashed line) in the chiral limit, considering nearest neighbor interlayer hopping.}
\label{fig1}
\end{center}
\end{figure}
%We first provide a lattice model of arbitrary alternating teisted graphene(ATMG) system. ATMG is a class of multilayer graphene conisists of three sequences of ordered stacking graphene. The sequences are labeled as '$M$', '$L$' and '$N$' sequence from bottom to top. The length of each sequence is M, L and N. In each sequence, the stacking chriality can be Bernal($ABA$), rhombohedral($ABC$) or mixed stacking. The '$N$'('$L$') graphene and the '$L$'('$M$') graphene are twisted by an angle $\theta$($-\theta$) as shown in Fig.~\ref{fig1}(a). Due to the small twist angle, the twsited graphene system forms a moir\'e supercell, and the moire lattice constance is $L_s\!=\!a/2\sin(\theta/2)$, where $a\!=\!2.46\,\angstrom$ is the atomic lattice constance of monolayer graphene.  A real space supercell indicates a relatively small moir\'e Brillouin zone shown in Fig.~\ref{fig1}(b). There are two inequivalent valley in the reciprocal space, each valley contains Driac fermions from each layers.
%\paragraph{Continuum model for ATMG. \textemdash}
\section{Continuum model for ATMG}
In this work we consider a class of alternating twisted multilayer graphene (ATMG), which consist of three sets of graphene multilayers with the number of layers denoted by $M$, $L$, $N$ respectively. The stacking sequence within each set of multilayers can be Bernal ($ABA$), rhombohedral ($ABC$), or a mixture of the two. These multilayers are stacked from bottom to up in the $M$-$L$-$N$ sequence, where the $N$ ($L$) layers and the $L$ ($M$) layers are twisted by an angle $\theta$ (-$\theta$) as schematically shown in Fig.~\ref{fig1}(a). Such a system forms a moir\'e pattern in real space with the moir\'e superlattice constant $L_s\!=\!a/(2\sin{\theta/2})$, where $a\!=\!2.46\,\angstrom$ is the graphene lattice constant. The corresponding moir\'e Brillouin zone is shown in Fig.~\ref{fig1}(b). Similar to twisted bilayer graphene (TBG), the low energy states of the ATMG system are contributed by those from the atomic $K$ and $K'$ valleys, which are approximately decoupled from each other at the non-interacting level for small twist angles. Thus, it is generally assumed that the system preserves valley charge conservation at small twist angles  \cite{BM}.    
Therefore, we generalize the Bistritzer-MacDonald continuum model \cite{BM} to describe the low-energy states of the ATMG system for each valley and each spin, assuming the states from the $K$ and $K'$ valleys are completely decoupled. The continuum model for valley $\mu$ ($\mu=\mp$ for $K$ and $K'$ valleys) is expressed as
%Having the lattice model of ATMG, we construct a continuum model to describe the low-energy electronic structure of ATMG. It can be constructed based on the Bistritzer-MacDonald continuum model\cite{BM} and  twist double bilayer graphene continuum model\cite{jpliu-tmg-prx}. In this model, we treat low energy states from two different atomic valleys as completely decoupled. For each valley, we contain Dirac fermions from each layers and the coupling between them. We only consider the interlayer hopping for two neighboring layers. For untwisted layers, only nearest neighbor hopping are contained. For layers with a twist between them, they are coupled through a moir\'e potential:
%
 \begin{align}
%\begin{split}
&H^{\mu}_{ATMG} 
= \left(\begin{array}{ccc}
H^{\mu}_{N} & \mathbb{U}_{\mu}^{\dagger}e^{-i\mu\Delta\mathbf{K}\cdot\mathbf{r}} & 0\\
\mathbb{U}_{\mu}e^{i\mu\Delta\mathbf{K}\cdot\mathbf{r}} & H^{\mu}_{L} & \mathbb{U}_{\mu}e^{i\mu\Delta\mathbf{K}\cdot\mathbf{r}}\\
0 &  \mathbb{U}_{\mu}^{\dagger}e^{-i\mu\Delta\mathbf{K}\cdot\mathbf{r}} & H^{\mu}_{M}
\end{array}\right)
%\end{split}
\label{eq:H-atmg}
\end{align}
where $H^{\mu}_{N}$, $H^{\mu}_{L}$ and $H^{\mu}_{M}$ denote the $\mathbf{k}\!\cdot\!\mathbf{p}$ Hamiltonians of the untwisted graphene multilayers, which consist of the Dirac fermions of each monolayer graphene and the interlayer hopping terms. $\mathbb{U}_{\mu}e^{i\mu\Delta\mathbf{K}\cdot\mathbf{r}}$ stands for the moir\'e potential term for valley $\mu$, which arises from the mutual twist between two sets of adjacent multilayers. $\Delta\mathbf{K}=(0,4\pi/(3L_s))$ is a vector characterizing the shift of Dirac points due to the twist.
The details of the continuum Hamiltonian Eq.~(\ref{eq:H-atmg}) are presented in Supplementary Information. 

%We provide band structures of $A-A-A$ and $A-ABA-A$ in Fig.~\ref{fig1}(c)(d). There are one pair of flat bands co-exists with a Dirac cone(per valley per spin) for $A-A-A$, while there are two pairs of flat bands for each valley and spin of $A-ABA-A$.

%%%%%%%%%%%%%%%%%%%%%%%%%%%%%%%%%%%%%%%%%%%%%%%%%%%%%%
%%%%%%%%%%%%%    partition rules   %%%%%%%%%%%%%%%%%%%%%%%%%%%%%%%%
%%%%%%%%%%%%%%%%%%%%%%%%%%%%%%%%%%%%%%%%%%%%%%%%%%%%%%
\section{Generic partition rules} 
%We can depict arbitrary ATMG systems with Eq.~(\ref{eq:H-atmg}). We figure out that ATMG systems can be roughly divided into three types based on their low energy band structure: one pair of flat bands with large band width(per valley per spin), one pair of flat bands co-exists with Dirac cone(quadratic bands) and $E(\mathbf{k})\sim k^{J}$($J$ is positive integer) as shown in Fig.~\ref{fig1}(c), two pairs of falt bands co-exist with $E(\mathbf{k})\sim k^{J}$($J$ is positive integer) as shown in Fig~\ref{fig1}(d). 

One can obtain various types of low-energy band structures from the Hamiltonian given by Eq.~(\ref{eq:H-atmg}). A careful study reveals that the ATMG systems can be roughly divided into three types based on their low-energy band dispersion: for type (\i) there  is only one pair of flat bands for each valley and spin, which is similar to TBG; for type (\ii) there is one pair of flat bands co-existing with some  low-energy bands characterized by the dispersion $E(\mathbf{k})\sim k^{J}$($J$ is positive integer); and for type (\iii) there are two pairs of flat bands. 
In Fig.~\ref{fig1}(c)-(d) we show the band structures of two typical ATMG systems with $A-A-A$ and $A-ABA-A$ stacking, where the solid and dashed lines denote energy bands from the $K$ and $K'$ valleys respectively in chiral limit, i.e the intrasublattice coupling between twisted layers are zero. For the $A-A-A$ system, there is  one pair of flat bands co-existing with a Dirac cone (per valley per spin), which can be categorized as type (\ii) ATMG; while there are two pairs of flat bands for each valley and spin for the $A-ABA-A$ system, which is the simplest example of type (\iii) ATMG. In what follows we will explain the origin of such intriguing low-energy dispersion and derive partition rules for generic ATMG systems.

%The fruitful low energy band structures of ATMG may have some rules. For a better understanding of the band structure of ATMG, we can treat it as ordered untwisted layers and neighboring layers with a twist angle. The untwisted layer in ATMG canbe decomposed into tensor product of each segment by a pertubation approach\cite{MacDonaldmultilayer}, which treats interlayer hopping chain as zero modes and intralayer Dirac cones as pertubation. In previous work, the author provided a partition rule for twist multilayer graphene(TMG): the bands of each segments are describled as pseudospin \cite{tmgdecomposition}:
%
To better illustrate the origin of these intriguing low-energy dispersion, we first consider the chiral limit in which all the intrasublattice couplings are turned off. Within the chiral limit, we first analyze the low-energy states of the untwisted multilayers, then discuss the effects of the moir\'e potentials at the twisted interfaces. It has been proposed that an untwisted multilayer graphene with arbitrary stacking sequence and with the total number of layers $N$ can be decomposed into $S_N$ chiral segments, within each of which the stacking chirality is unchanged. Then the low-energy states contributed by the $i$th chiral segment with the number of layers $J_i$ consists of a chiral doublet described by the following effective Hamiltonian
\begin{equation}
H_{J_{i}}^{\mu}(\textbf{k})\propto k^{J_{i}}[\cos(J_i \phi_{\textbf{k}})\sigma_{x}\pm\sin(J_i \phi_{\textbf{k}})\sigma_{y}]
\label{eq:doublet}
\end{equation}
where $\tan{\phi_{\mathbf{k}}}=k_y/k_x$, and $\sigma_{x,y}$ denote Pauli matrices in the sublattice space. Then the low-energy Hamiltonian of the untwisted $N$ layers from valley $\mu$ can be written as a direct sum of those of the $N_D$ chiral segments:
\begin{equation}
H^{\mu}_{N}\approx H^{\mu}_{J_1}\oplus H^{\mu}_{J_2}... \oplus H^{\mu}_{J_{S_N}}\;.
\label{eq:segment}
\end{equation} 
Each  segment contributes to a chiral doublet with the dispersion $E\sim k^{J_{i}}$ around $\textbf{K}^{\mu}$ point.  
Then we consider the chirally decomposed $N$ layers are stacked with the other $M$ layers and are twisted by angle $\theta$. The moir\'e potential at the interface would couple  the topmost  chiral segment of the $N$ with the bottom-most segment of the $M$ layers, giving rise to a pair of flat bands for each spin and each valley. These flat bands would  co-exist with the dispersive chiral doublets contributed by the remaining chiral segments (if any) of the two sets of multilayers\cite{tmgdecomposition}. The ATMG system introduces additional complexity due to the additional multilayers ($L$ layers) and the additional twist. It turns out that the situations with the number of middle layers $L\!=\!1$ and $L\!>\!1$ need to be treated separately.

%which turns out a pair of bands with the dispersion relation $E\sim k^{J_{i}}$ at $\textbf{K}$($\textbf{K}^{\prime}$) point. The difference between TMG and ATMG is the additional alternating twist. For L$>$1, two moir\'e potential are not coupled directly; while for L=1, the dispersion relation may be different.

To evaluate the difference between ATMG systems with $L\!=\!1$ and $L\!>\!1$, we first consider alternating twisted trilayer graphene (TTG), i.e $M\!=\!L\!=\!N\!=\!1$. We will   show that for TTG, in which three alternating twisted layers are coupled together, can be decoupled into a TBG-like Hamiltonian and a free Dirac-fermion Hamiltonian. 
%To start with, the constant wavevectors $\textbf{K}_l$ in $\mathbb{U}$ can be gauged: $\psi^{\prime}_{ls}(\textbf{r})=\psi_{ls}(\textbf{r})e^{i\textbf{K}_{l}\cdot\textbf{r}}$, where $l$ denotes layer, $s$ denotes AB sublattice\cite{liu-pll}. The interlayer moir\'e  potential can be expanded to the first order of $r/L_s$, then the Hamiltonian can be rewritten into : $\tilde{H}=-\hbar v_{F} \mathbbm{1}_{3\times 3}\otimes \bm{\sigma} + \frac{2\pi}{L_{s}}u^{\prime}_{0}(r_{x},r_{y})\tilde{\tau_{y}}\otimes \bm{\sigma} + 3u_{0}\tilde{\tau_{x}}\otimes \mathbbm{1}_{2 \times 2}$. %, with $\tilde{\tau_y}$ defined in the layer subspace:
%
%%%%%%%%%%%%%%%
%\iffalse
%\begin{align}
%\begin{split}
%\tilde{\tau_{y}}=\left(\begin{array}{ccc}
%				0 & i & 0\\
%				-i &0 & -i\\
%				0& i & 0
%\end{array}\right)
%\end{split}
%\end{align}
%\fi
%%%%%%%%%%%%%%%%%
%Transform $\tilde{H}$ to diagonalize $\tilde{\tau_{y}}$, we get a decoupled free fermion and two coupled fermion under pseudo magnetic field. Apply this transformation to the unexpanded continuum model, we have a similar form. Then apply the inversr transform to the two coupled fermion, the total transformation is:
For the sake of convenience, we first apply a gauge transformation to the basis functions of the ATMG system:  
\begin{equation}
\widetilde{\psi}_{ls,\mathbf{k}}^{\mu}(\textbf{r})=\psi_{ls,\mathbf{k}}^{\mu}(\textbf{r})e^{i\textbf{K}_{l}^{\mu}\cdot\textbf{r}}
\label{eq:gauge}
\end{equation}
where $\mathbf{K}^{\mu}_{l}$ denotes the Dirac point of valley $\mu$ and layer $l$, and $s$ is the sublattice index. Such a gauge transformation would  remove the  phase factor $e^{i\pm\mu\Delta\mathbf{K}\cdot\mathbf{r}}$  in the moir\'e potential term, and would move the Dirac points of the different twisted layers to the same origin.
Then we apply a  unitary transformation to the alternating TTG Hamiltonian, $\tilde{H}^{\mu}_{TTG}=W^{\dagger}H^{\mu}_{TTG}W$. The Hamiltonian after  the unitary transformation $\tilde{H}^{K}_{TTG}$ is expressed as
 \begin{align}
\begin{split}
\tilde{H}^{\mu}_{TTG} 
= \left(\begin{array}{ccc}
h^{\mu}(\mathbf{k}) &0 & \sqrt{2}\,\mathbb{U}_{\mu}\\
0 & h^{\mu}(\mathbf{k}) & 0\\
\sqrt{2}\,\mathbb{U}_{\mu}^{\dagger} &  0 & h^{\mu}(\mathbf{k})
\end{array}\right)\;,
\end{split}
\label{eq:H-ttg}
\end{align}
where $h^{\mu}(\mathbf{k})=-\hbar v_{F}\textbf{k}\cdot\bm{\sigma}_{\mu}$, with the Pauli matrices $\bm{\sigma}_{\mu}=(\mu\sigma_x,\sigma_y)$ defined in the sublattice space, and the unitary transformation matrix $W$ is expressed as
\begin{align}
\begin{split}
W=\left(\begin{array}{ccc}
				0 & -\frac{1}{\sqrt{2}} & \frac{1}{\sqrt{2}}\\
				1 &0 & 0\\
				0& \frac{1}{\sqrt{2}} & \frac{1}{\sqrt{2}}
\end{array}\right)\;.
\end{split}
\label{eq:transform}
\end{align}
From Eq.~(\ref{eq:H-ttg}) it is immediately seen that the total Hamiltonian of alternating twisted trilayer graphene consists of a TBG-like part with the moir\'e potential rescaled by $\sqrt{2}$ and a free Dirac fermion part. The TBG part and Dirac fermion part are completely decoupled from each other, which  contribute to one pair of flat bands co-existing with a Dirac cone as shown in Fig.~\ref{fig1}(c). Moreover, since the magic angle is determined by the ratio between the intersublattice component of the moir\'e potential and the Fermi velocity,  the re-scaled moir\'e potential in Eq.~(\ref{eq:H-ttg}) implies that the magic angle for the TTG system is rescaled by the same factor, i.e., the new magic angle should be $\sqrt{2}\times 1.05^\circ\approx 1.5^{\circ}$.

For $L\!=\!1$ but $M, N\!>\! 1$, one can apply the chiral decomposition rule as discussed in Eqs.~(\ref{eq:doublet})-(\ref{eq:segment}) to the $M$ layers and $N$ layers. The topmost chiral segment from the $M$ layer and the bottom-most segment from the $N$ layers are coupled with the $L\!=\!1$ middle layer through the moir\'e potentials, contributing to one pair of flat bands co-existing with either a  Dirac cone or a pair of quadratic bands. The remaining chiral segments (if any) in the $N$ layers and $M$ layers would contribute to additional $E(\mathbf{k})\sim k^{J}$ dispersive bands.  On the other hand, when $L\!>\! 1$, one need to apply the chiral decomposition rule to the $L$ multilayers as well, and carefully study how the chiral doublets contributed by the $M$, $L$, and $N$ layers are coupled to each other through the moir\'e potentials at the two twisted interfaces.
% then we have one free fermion decoupled with a TBG-like term:
%
% \begin{align}
%\begin{split}
%\tilde{H}^{K}_{TTG}=W^{\dagger}H^{K}_{TTG}W = \left(\begin{array}{ccc}
%h &0 & \sqrt{2}\cdot\mathbb{U}\\
%0 & h & 0\\
%\sqrt{2}\cdot\mathbb{U}^{\dagger} &  0 & h
%\end{array}\right)
%\end{split}
%\label{eq:total-transformation}
%\end{align}
%where $h=-\hbar v_{F}\textbf{k}\cdot\bm{\sigma}$. This model indicates that three Dirac cones coupled through the moir\'e potential can be regarded as a decoupled Dirac cone co-exists with a TBG with a $\sqrt{2}$ enlarged interlayer potential. The enlarged interlayer potential explains the reason why Fermi velocity vanishes at $1.5^\circ\approx\sqrt{2}\times 1.05^\circ$.

%As we mentioned above, for untwisted layer, we can divide them into segments with $E\sim k^{J_{i}}$ at $\textbf{K}$($\textbf{K}^{\prime}$); for twisted layers, they can be decomposed to a TMG-like form. Then we come up with the partition rules for ATMG:
After a comprehensive theoretical analysis based on a simplified $\mathbf{k}\!\cdot\!\mathbf{p}$ model approach (to be explained in the following section), we have derived a set of generic partition rules describing the low-energy band structures of  ATMG systems in the chiral limit. First, the $M$, $L$, and $N$ multilayers are divided into $S_{M}$, $S_{L}$, and $S_N$ chiral segments (see Eqs.~(\ref{eq:doublet})-(\ref{eq:segment})), and the number of layers of the $i$th segment, say, in $N$ multilayer is denoted as $J_{N, i}$ ($i=1,..., S_N$). We also need to keep the chiral segments that are closest to the twisted interfaces to be as long as possible, i.e., we need to make a choice of chiral decomposition to make $J_{M,S_{M}}$, $J_{L,1}$, $J_{L,S_{L}}$ and $J_{N,1}$ as large as possible. Based on the above choice of chiral segments, we reach the following partition rules for the low-energy dispersion in the chiral limit:
%Divide the sequence into $D_{M}+D_{L}+D_{N}$ segments based on their stacking chirality. $J_{a,b}$, where $a=M,L,N, b=D_{j}$ is the length of the j-th segments in '$a$' sequence. We have $a=\sum_{j=D_{1}}^{D_{a}}J_{a,j}, a=M,L,N$. Keep the segments containing the twisted layers as long as possible(to be specific, keep $J_{M,D_{M}}, J_{L,D_{1}}, J_{L,D_{L}} and J_{N,D_{1}}$ as large as possible).

(a) For $L=1$: the $J_{M,S_{M}}$, $J_{L,1}$ and $J_{N,1}$ chiral segments are coupled through the  moir\'e potential generated by the alternating twisted structure. When the stacking chirality of $J_{M,S_{M}}$ and $J_{N,1}$ are the same, there are one pair of flat bands and one Dirac cone co-existing near $\textbf{K}_{\mu}$ point (per spin per valley); while if $J_{M,S_{M}}$ and $J_{N,1}$ have opposite stacking chiralities, there are one pair of flat bands and one pair of quadratic bands  co-existing near $\textbf{K}_{\mu}$ point. The remaining chiral segments in the $M$ ($N$) multilayers would contribute to additional chiral doublets with the dispersion $E(\mathbf{k})\!\sim\! k^{J_{M,i}}$ ($E(\mathbf{k})\!\sim\! k^{J_{N,i}}$)  near  $\mathbf{K}_{\mu}$ ($\mathbf{K}_{\mu}'$) point.
%For the rema chiral segments, they act as Eq.~(\ref{segments}). Each segment turns out two bands with the dispersion to the power of $J_{i}$ near $\textbf{K}$($\textbf{K}^{\prime}$) point at CNP. 

(b) For $L>$1: if the $L$ multilayer can be divided into more than one chiral segments ($S_L>1$), there are two pairs of flat bands (double flat bands) around CNP; while if the $L$ multilayer is in the chiral  (or rhombohedral) stacking sequence ($S_L=1$), there is only one pair of flat bands. %in the chiral limit and the bandwidth will be enhanced with finite $w_{AA}$. 
When $S_L\!>\!1$, the remaining chiral segment $\{J_{L,i},\; 2\leq i\leq S_L-1\}$ that are not coupled with the $M$ and $N$ multilayers would contribute to dispersive bands $E(\mathbf{k})\sim k^{J_{L,i}}$ around CNP. Similarly, the remaining chiral segments  $\{J_{M,i}, \;1\leq i\leq S_M-1\}$ ($\{J_{N,i},\; 2\leq i\leq S_N$\}) from the $M$ ($N$) multilayer that are not coupled with the middle $L$ multilayer would contribute to the low-energy dispersive bands with $E(\mathbf{k})\sim k^{J_{M,i}}$ ($E(\mathbf{k})\sim k^{J_{N,i}}$).
%Apply partition rules to each sequence, segments $J_{L,i}, 1<i<D_{L}$ turns out bands touching at $\textbf{K}$($\textbf{K}^{\prime}$) point;  segments $J_{M,i}, i<D_{M}$ and $J_{N,i}, 1<i$ contribute bands touching at $\textbf{K}^{\prime}$($\textbf{K}$) point as Eq.~(\ref{segments}).

%[JPL: STOP HERE!!!]

In Table~\ref{tab:partitioning}, we illustrate some ATMG systems as typical cases and apply the partition rules described above to these systems to characterize their low energy band structures, where the notation $(m,n)$ means that there are $m$ pairs of bands with dispersion $E(\textbf{k})\sim\textbf{k}^{n}$ around $K_s$ or $K_s'$ points.  For example, for $A$-$A$-$ABAC$ system, it can be divided into two parts including the alternating twisted layers $A$-$A$-$AB$  and  untwisted layers $AC$. The twisted layers would contribute one pair of flat bands co-existing with a Dirac cone around $K_{\mu}$ point, while the untwisted layers give rise to a pair of quadratic bands($E(\textbf{k})\sim\textbf{k}^{2}$) centered at $K_{\mu}$ point. 

\begin{table}[]\scriptsize
    \centering
    \caption{Typical cases for generic partition rules. The band structures are measured with full continuum model considering nearest neighbor for untwisted bands. We label $m$ bands with dispersion $E(\textbf{k})\sim\textbf{k}^{n}$ at $\textbf{K}_{\mu}(\textbf{K}^{\prime}_{\mu})$ as (m,n). }
    \begin{tabular}{l|c|c|c}
    \hline\hline
     partitioning   & number of & bands at $\textbf{K}$ & bands at $\textbf{K}^{\prime}$ \\
      &falt bands & & \\
      \hline
       A-A-A & 2 & (1,1) & 0 \\
       A-A-AB+AC  & 2 & (1,1),(1,2) &0 \\
       AB-A-BA  & 2 & (1,1) & 0 \\
       AB-A-AB  & 2 & (1,2) & 0 \\
       A-AB+A-A & 4 & 0 & 0 \\
       A-ABC-A  & 2 & / & / \\
       A-AB+ABC-A  & 4 & 0 & 0 \\
       A+BA-AB+A+BA-AB+A &4 & (1,1) & (2,1) \\
       \hline\hline
    \end{tabular}
    \label{tab:partitioning}
\end{table}

%For the low-energy physics, we consturct simplified$\textbf{k}\cdot \textbf{p}$ model to describe energy bands near the moir\'e $K$ or $K^{\prime}$ point at CNP. For each TBG-like term, we take the zero mode solution for flat bands, and keep other layers as free fermion.  Then we only consider the first nearset neighbor hopping between untwisted layers. 

\section{Simplified $\textbf{k}\cdot\textbf{p}$ model} 
%We can obtain the low energy band structure of arbitrary stacking ATMG through both continuum model and generic partition rules. In this section, we will show that the full continuum Hamiltonion has redundancy for the low energy band structure and we can simplify it with a effective Hamiltonian, named simplified $\textbf{k}\cdot\textbf{p}$ model.  
%According to the generic partition rules, we can treat an ATMG system as twisted segments and untwisted ones. The twisted segments contribute flat bands and the untwisted segments turn out dispersive bands with $E(\textbf{k})\sim\textbf{k}^{J_{i}}$. In the perspective of low energy bands dispersion, the dominate terms are the zero modes from twisted segments and the dispersion near band touching point. % from untwisted segments. 
The partition rules presented above can be derived using a simplified $\mathbf{k}\cdot\mathbf{p}$ model approach. In this approach, we write a $\mathbf{k}\cdot\mathbf{p}$ model within the moir\'e Brillouin zone by expanding the flat bands and the Dirac cones around the moir\'e $K_s$ or $K_s'$ points including the coupling terms between them. In the chiral limit both the flat bands and the Dirac cone can be solved exactly \cite{ashvin-analytical-tbg}, then we can analytically construct a greatly simplified  $\mathbf{k}\cdot\mathbf{p}$ model in the basis of the zero modes (flat-band wavefunctions) and the Dirac fermions, and solve it exactly. From the analytic solutions of the simplified $\mathbf{k}\cdot\mathbf{p}$ model, we derive the  partition rules for generic ATMG systems presented above. Such an approach can capture the essential low-energy physics, while neglecting the irrelevant high-energy bands obtained from a direct numerical diagonalization of the original continuum Hamiltonian.

To construct such a simplified $\textbf{k}\cdot\textbf{p}$ model for a generic ATMG system (in the chiral limit), we should first  find  proper unitary transformations to the original continuum Hamiltonian to decompose it into a form consisting of  a TBG-like continuum Hamiltonian and free Dirac fermions, e.g., as illustrated in Eq.~(\ref{eq:H-ttg}). %Taking use of the exact zero-mode solutions for the TBG-like Hamiltonian at the (renormalized) magic-angle \cite{ashvin-analytical-tbg}, then expanding the Dirac cones around the moir\'e Dirac points $K_s$ or $K_s'$ within the moir\'e Brillouin zone, we immediately obtain a simplified  $\mathbf{k}\cdot\mathbf{p}$ model in the basis of the  flat-band wave functions and the low-energy Dirac-fermion wave functions, with all the irrelevant high-energy bands neglected. 
To be specific, for any ATMG system with $L\!=\!1$, we can apply the unitary transformation given by Eq.~(\ref{eq:transform}) to the three alternating twisted layers, and keep other layers unchanged. The corresponding layer-mixed basis functions are: 
\begin{align}
&\bar{\psi}_{\alpha s,\mathbf{k}}^{\mu}(\textbf{r})=\widetilde{\psi}_{Ls,\mathbf{k}}^{\mu}(\textbf{r})\;\nn
&\bar{\psi}_{\beta s,\mathbf{k}}^{\mu}(\textbf{r})=\frac{-1}{\sqrt{2}}\widetilde{\psi}_{Ms,\mathbf{k}}^{\mu}(\textbf{r})+\frac{1}{\sqrt{2}}\widetilde{\psi}_{Ns,\mathbf{k}}^{\mu}(\textbf{r})\;\nn
&\bar{\psi}_{\gamma s,\mathbf{k}}^{\mu}(\textbf{r})=\frac{1}{\sqrt{2}}\widetilde{\psi}_{Ms,\mathbf{k}}^{\mu}(\textbf{r})+\frac{1}{\sqrt{2}}\widetilde{\psi}_{Ns,\mathbf{k}}^{\mu}(\textbf{r})
\label{eq:unitary-transform}
\end{align}
where $\alpha$, $\beta$, $\gamma$ are the three mix-layer indices marking the basis functions after the unitary transformation, $\widetilde{\psi}_{l s,\mathbf{k}}^{\mu}(\textbf{r})$ with $l=M, L, N$ denote the basis wave functions (after the gauge transformation of Eq.~(\ref{eq:gauge})) of the three alternating twisted layers in the $M$, $L$, and $N$ sequence respectively, and $s=A, B$ is the sublattice index. 
 After such a transformation, the continuum Hamiltonian consists of a TBG-like part and a free Dirac fermion (as shown in Eq.~(\ref{eq:H-ttg}) contributed by the three alternating twisted layers, with the terms from the other layers being unchanged. One can solve for the zero-mode solutions for the  TBG-like part at the renormalized magic angle $\sqrt{2}\times 1.05^{\circ}$, and expand all the Dirac cones around the Dirac points within the moir\'e Brillouin zone, then eventually obtain a $\mathbf{k}\cdot\mathbf{p}$ model in a greatly simplified form.
For L$>$1, above transformation is unnecessary, since there are two pairs of twisted layers giving rise to two TBG-like terms. For each of the TBG-like terms, we can obtain  the zero-mode solution of magic-angle TBG in the chiral limit. The analytical wave functions for the zero modes in magic-angle TBG  in the chiral limit are expressed as \cite{ashvin-analytical-tbg}: 
\begin{align}
\Psi_{lA,\textbf{k}}(\textbf{r})=f_{\textbf{k}}(z)\bar{\psi}_{l A,\textbf{K}}(\textbf{r}),\;\nn
\Psi_{lB,\textbf{k}}(\textbf{r})=f^{*}_{\textbf{k}}(-z)\bar{\psi}_{l B,\textbf{K}}(\textbf{r})
\end{align}
where $l=\alpha,\gamma$ refers to the two mix-layer indices as defined in Eq.~(\ref{eq:unitary-transform})
, $\Psi_{l s,\textbf{K}}(\textbf{r})$ refers to the $s$ ($s=A, B$) sublattice component of the zero-mode solution at the Dirac point $\mathbf{K}$ \cite{ashvin-analytical-tbg}. %$f_{\textbf{k}}(x+iy)=\vartheta_{(\textbf{ka}_{1}/2\pi)-\frac{1}{6},\frac{1}{6}-(\textbf{ka}_{2}/2\pi)}((x+iy)/(a_{1,x}+ia_{1,y})\mid e^{i2\pi/3})/\vartheta_{-\frac{1}{6},\frac{1}{6}}((x+iy)/(a_{1,x}+ia_{1,y})\mid e^{i2\pi/3})$, where $\vartheta_{a,b}(x+iy\mid\tau)=\sum_{n=-infty}^{\infty}e^{i\pi\tau(n+a)^{2}}e^{2\pi i(n+a)(n+b)}$. 
$f_{\textbf{k}}(z)=\vartheta_{a_{1},b_{1}}(z)/\vartheta_{a_{2},b_{2}}(z)$, where $\vartheta_{a,b}(z)$ is the theta function defined in the previous work\cite{ashvin-analytical-tbg}. 
%The analytic zero mode wavefunction has good quantum number on both x and y axis in reciprocal space. 
For the untwisted layers, we take the $\textbf{k}\cdot\textbf{p}$ Hamiltonians of the Dirac fermions and expand them around the Dirac points within the moir\'e Brillouin zone. The coupling between the zero modes from the twisted layers and the Dirac fermions from the untwisted layers can be evaluated by re-expressing the original interlayer hopping matrix in the basis of the zero-mode wavefunctions and the free Dirac-fermion states. %, the details of which are presented in Supplementary Information.

%We take $AB$-$A$-$AB$ and $AB$-$A$-$BA$ as an example, which can explain the dispersion relation in partition rules (a). 
We take the $AB$-$A$-$AB$ and $AB$-$A$-$BA$ stacked ATMG systems as two typical examples to illustrate how the partition rule (a) works.
Following the procedures described above, we obtain the simplified $\textbf{k}\cdot\textbf{p}$ Hamiltonian for $AB$-$A$-$AB$ ($AB$-$A$-$BA$) stacked ATMG system: 
\begin {align}
\begin{split}
H^{\mu}_{\textbf{k}\cdot\textbf{p}}&=\left(\begin{array}{ccccc}
				h^{\mu}(\textbf{k}) & 0 &-h_{+}& \tilde{h}_{+}&0\\
				0 &0 &0& 0&0\\
				-h_{-}& 0& h^{\mu}(\textbf{k}) & 0 &h_{+(-)} \\
				\tilde{h}_{-}&0&0& 0 &\tilde{h}_{+(-)}\\
				0&0&h_{-(+)}&\tilde{h}_{-(+)}&h^{\mu}(\textbf{k})
\end{array}\right)\;,
\end{split}
\label{eq:AB-A-ABkp}
\end{align}
where 
\begin{align}
h_{+}&=\left(\begin{array}{cc}
				0 & 0 \\
				t_{\perp} &0
\end{array}\right)\;,\tilde{h}_{+}=\left(\begin{array}{cc}
				0 & 0 \\
				\tilde{t}_{\perp} &0
\end{array}\right)\;
\label{eq:tildeh}
\end{align}
$h^{\mu}(\textbf{k})=-\hbar v_{F}\textbf{k}\cdot\bm{\sigma}^{\mu}$, and $h_{-}=h_{-}^{\dagger}$ for different stacking chirality. The ``$+(-)$" sign in  $\tilde{h}_{+(-)}$ in Eq.~(\ref{eq:AB-A-ABkp}) applies to the $AB$-$A$-$AB$ ($AB$-$A$-$BA$) stacking. $t_{\perp}$ is the interlayer intersublattice hopping parameter. In the chiral limit we only consider nearest neighbor interlayer hopping term, and $\tilde{t}_{\perp}$ is the renormalized hopping parameter. We refer the readers to Supplementary Information(Appendix II) for more details about the simplified $\textbf{k}\cdot\textbf{p}$ model. %

\begin{figure}[b]
\begin{center}
    \includegraphics[width=8.5cm]{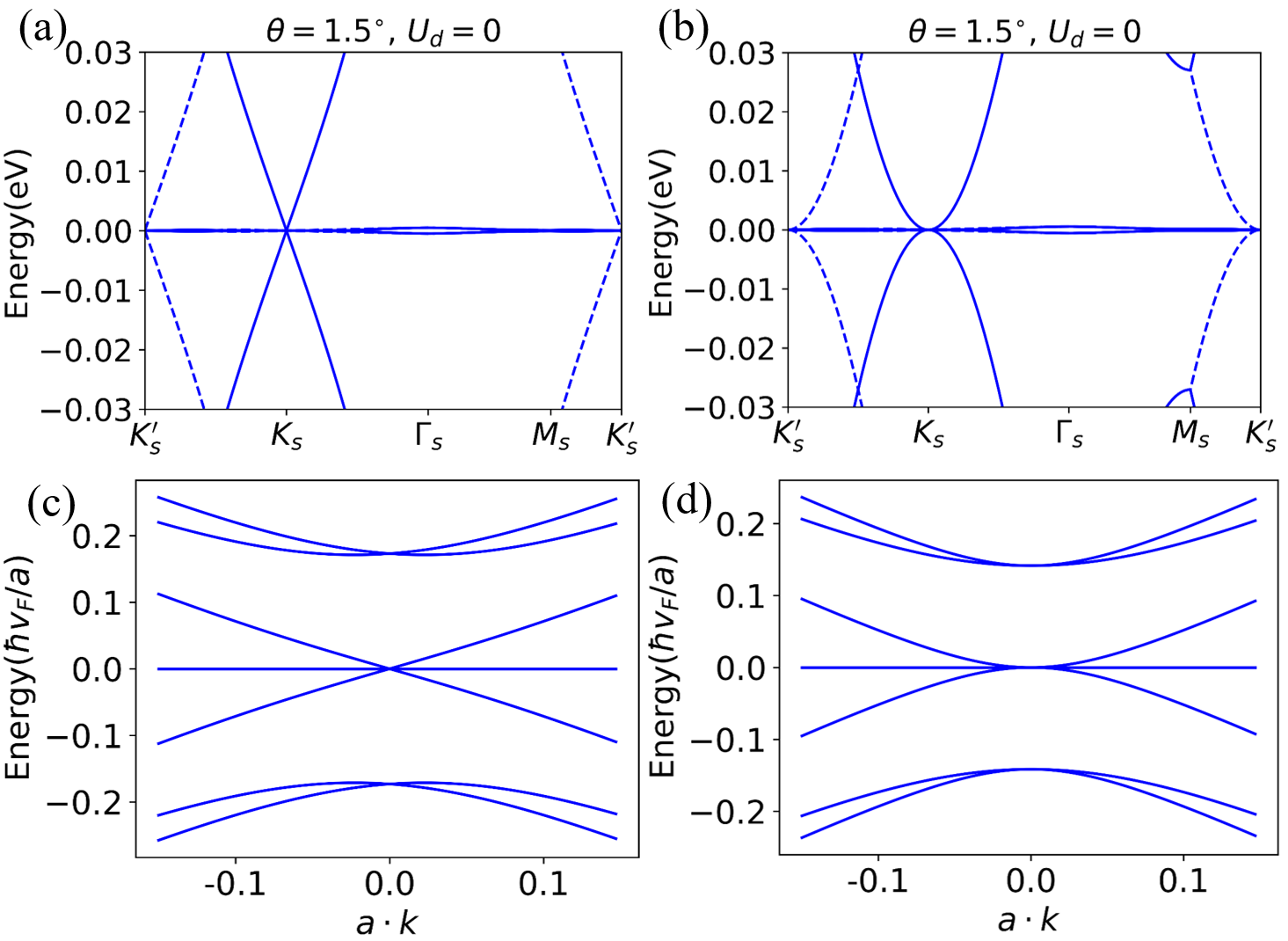}
\caption{
The band structures of $AB-A-AB$ for $K$ valley(solid line) and $K^{\prime}$ valley(dashed line) obtained from (a) continuum model and (c) simplified $\textbf{k}\cdot\textbf{p}$; $AB-A-BA$ for K valley obtained from (b) continuum model and (d) simplified $\textbf{k}\cdot\textbf{p}$ model. The continuum models are constructed including nearest neighbor interlayer hopping and intersublattice moir\'e potential.}
\label{fig2}
\end{center}
\end{figure}

%We provide the band structures of above two ATMG in continuum Hamiltonian and simplified $\textbf{k}\cdot\textbf{p}$ Hamiltonian in Fig.~(\ref{fig2}). We obtain that: for $AB-A-AB$, there exists two zero modes and one Dirac cone; for $AB-A-BA$, there exists two zero modes and a pair of quadratic bands touching at CNP. This effect caused by stacking chirality near the twist layers remains correct for the continuum model Eq.~(\ref{eq:H-atmg}). We can describe it with a 10$\times$10 matrix instead of a full continuum Hamiltonian. 
We present the band structures of the above two ATMG systems in Fig.~(\ref{fig2}). In particular, in Fig.~\ref{fig2}(a)-(b) we show the band structures for $AB$-$A$-$AB$ and $AB$-$A$-$BA$ systems obtained from direct numerical diagonalizations of the full continuum Hamiltonian in chiral limit, while in Fig.~\ref{fig2}(c)-(d) the band structures are calculated by the  $10\times 10$ simplified $\mathbf{k}\cdot\mathbf{p}$ model expanded around either $K_s$ or $K_s'$ point as given by Eq.~(\ref{eq:AB-A-ABkp}). For the  $AB$-$A$-$AB$ system there are two flat bands co-existing with a Dirac cone (Fig.~\ref{fig2}(a) and (c)), while for the  $AB$-$A$-$BA$ system there are two flat bands and one pair of quadratic band touching around $K_s$ ($K_s$') for $K$ ($K'$) valley. We see that the $10\times 10$ simplified $\mathbf{k}\cdot\mathbf{p}$ model can fully capture the essential low-energy physics, which fulfills the partition rules introduced above.
%fig2

%%%%%%%%%%%%%%%%%%%%%%%%%%%%%%%%%%%%%%%%%%%%%%%%%%%%%%
%%%%%%%%%%%%%%%%   double flat bands  %%%%%%%%%%%%%%%%%%%%%%%%%%%%
%%%%%%%%%%%%%%%%%%%%%%%%%%%%%%%%%%%%%%%%%%%%%%%%%%%%%%
\section{Origin of double flat bands} Some of the ATMG systems may have two pairs of  flat bands (double flat bands) for each valley and spin. 
According to the partition rules, the double flat bands can only occur when $S_{L}>$1. %The origin of double flat bands can also be described by the simplified $\textbf{k}\cdot \textbf{p}$ model. 
%First we consider $A-ABA-A$(with double flat bands) and $A-ABC-A$(one pair of flat bands) as simple cases. %The simplified $\textbf{k}\cdot \textbf{p}$ Hamiltonian for $A-ABA-A$: 
To illustrate the origin of the double flat bands, we first study two ATMG systems for $L\!=\!3$ with $A$-$ABA$-$A$ (with double flat bands) and $A$-$ABC$-$A$ (with one pair of flat bands) stacking based on the simplified $\mathbf{k}\cdot\mathbf{p}$ model.  
%
%It is straightford to check that there are four zero modes for the simplified $\textbf{k}\cdot\textbf{p}$ Hamiltonian of $A-ABA-A$, which turns out double flat bands , while there are only two zero modes for that of $A-ABC-A$.
To be specific, the simplified $\textbf{k}\cdot\textbf{p}$ Hamiltonian of the $A$-$ABA$-$A$ system can be written as a direct sum of a $6\times6$ matrix and two zero modes
\begin {align}
\begin{split}
H^{\textbf{k}\cdot\textbf{p}}_{A-ABA-A}=\left(\begin{array}{ccc}
				0& \tilde{h}_{+}& 0  \\
				\tilde{h}_{-}&h(\textbf{k})&\tilde{h}_{-}\\
				0&\tilde{h}_{+}&0
\end{array}\right) \oplus \left(\begin{array}{cc}
				0 &0 \\
				0&0
\end{array}\right)\;,
\end{split}
\label{eq:A-ABA-A}
\end{align}
where the $6\times 6$ matrix consists of two zero modes coupled with a free Dirac fermion. Then it is straightforward to check that the eigenvalues of this $6\times 6$ matrix has two  zero modes, so that the entire system has four zero modes, leading to double flat bands. 
For $A$-$ABC$-$A$,  the simplified $\textbf{k}\cdot\textbf{p}$ model can be written in a similar form
\begin {align}
\begin{split}
H_{A-ABC-A}=\left(\begin{array}{ccc}
				0& \tilde{h}_{+}& 0  \\
				\tilde{h}_{-}&h(\textbf{k})&\tilde{h}_{+}\\
				0&\tilde{h}_{-}&0
\end{array}\right) \oplus \left(\begin{array}{cc}
				0 &0 \\
				0&0
\end{array}\right)
\end{split}
\label{eq:A-ABC-A}
\end{align}
Due to the change of stacking chirality, the $6\times 6$ Hamiltonian consisting of two zero modes coupled with a free Dirac fermion no longer has zero-mode solution. Thus, there are only two zero modes for the $A$-$ABC$-$A$ system, contributing to one pair of flat bands per spin per valley.
%The former matrix has no zero modes, so that the system has only two zero modes.% Since the condition of simplified $\textbf{k}\cdot\textbf{p}$ model is under chiral limit, the band width of zeros modes is zero. For finite $w_{AA}$, the zero modes become more dispersive. 

%For finite $w_{AA}$, a simplified $\textbf{k}\cdot \textbf{p}$ model in pseudo Landau level approach can explain why four zero modes reamin for $A-ABA-A$ while no flat bands for $A-ABC-A$ without chiral limit. This is straightforward by taking higher pesudo Landau levels into account.

We can generalize the above argument for $L>3$ and $M,N>$1. According to the partition rules, the number of chiral segments in the $L$ multilayers have a dramatic influence on the number of flat bands, while the segments in the $M$ and $N$ multilayers do not. First, if the middle $L$ multilayers has a chiral stacking, i.e., $S_L\!=\!1$, then the system is equivalent to the $A$-$ABC$-$A$ system with a renormalized interlayer hopping parameter between untwisted layers $t^{\prime}_{\perp}$. 
%$(a)$ For chiral stacking in the middle  sequences, i.e $S_{L}=1$, it can be treated as $A-ABC-A$ with a renormalized interlayer hopping parameter between untwisted layers $t^{\prime}_{\perp}$. 
As a result, there is only one pair of flat bands (in the chiral limit) for the middle $L$ layers with chiral stacking. Second, if the $L$ multilayers can be decomposed into two segments($S_{L}\!=\!2$), the corresponding simplified $\mathbf{k}\cdot\mathbf{p}$ model is equivalent to that of the $A-ABA-A$ system with renormalized hopping parameters, which has double flat bands.  Finally,  for $S_{L}\!>\!2$: % we can 
we can argue that for alternating twist mutilayers, two pairs of moir\'e potentials are coupled together through several chiral segments within the $L$ multilayers, so that the more the segments are, the weaker the couple is.  As a result, we can treat them as nearly decoupled and there are four flat bands for the $S_{L}\!>\!2$ situation.   Therefore, if the middle $L$ multilayers has $S_L\!>\!1$, there would be double flat bands for each spin and valley degrees of freedom.

It is worthwhile to note that the simplified $\mathbf{k}\cdot\mathbf{p}$ model is constructed in the chiral limit neglecting all the intrasublattice couplings. Consequently, the partition rules derived from the  simplified $\mathbf{k}\cdot\mathbf{p}$ model and the above discussions about double flat bands are rigorous only in the chiral limit.  In a more realistic situation, one needs to include the intrasublattice component of the moir\'e potential and the further neighbor interlayer hoppings within the untwisted layers as in Eq.~(\ref{eq:fur}), which break chiral symmetry. 
With these additional coupling terms, the otherwise exactly flat bands at the magic angle may acquire nonzero bandwidths, and the $E(\mathbf{k})\!\sim\!k^J$ bands may have slightly modified dispersion. However, despite these perturbative changes, the main conclusions sketched by the partition rules  are unchanged.

It follows from the previous arguments that a mirror-symmetric  ATMG system with $L\!>\!1$  must satisfy the condition of $S_{L}\!>\!1$, thus  there must be two pairs of flat bands. % The eigenvalue of the double flat bands are $\pm1$. 
The double flat bands  can be classified  by the opposite mirror eigenvalues $\pm 1$ for ATMG with $m_z$ symmetry\cite{li-arxiv19}.  In Fig.~\ref{fig:131}(a) we present the band structures of the $A$-$ABA$-$A$ system including the intrasublattice moire potential and the further neighbor interlayer hopping, where the color coding indicates the weight projected onto the middle layer. First, we note that compared with the band structures in the chiral limit shown Fig.~\ref{fig1}(d),  in the realistic situation two of the four flat bands become more dispersive with the bandwidth $\sim\!25\,$meV, while the other pair of flat bands remain flat with very small bandwidth $\sim 10\,$meV. Second, we note that the weight of the middle layer for the pair of flat bands lower in energy with small bandwidth is vanishing, while the upper pair of flat bands with  relatively large bandwidth have significant contributions from the middle-layer states. This is because the two flat bands lower/upper in energy have mirror eigenvalues $\mp1$, and the Bloch states with $-1$ mirror eigenvalue must have zero contribution from the middle layer. In the presence of Coulomb interactions, the $m_z$ symmetry could be broken spontaneously at certain filling factors. 

%We provide this result in Fig.~\ref{fig:131}(a) for $A$-$ABA$-$A$ as an example. In order to reach a realistic situation, we take further neighbor interlayer hopping for untwisted layers into account\cite{jpliu-tmg-prx}. It dose not break the $m_z$ symmetry. We find that one pair of flat bands become dispersive by further neighbor hopping while another pair remains flat. 
%The $m_z$ eigenvalues are +1 for dispersive flat bands, while eigenvalues become -1 for bands remaining flat. %Besides, the layer distribution to each bands are different: the middle layer $L_2$ contribute exactly zero to bands with $m_z$ eigenvalue -1. This can be proved by applying $m_z$ operator to the wavefunction, and for the middle layer, wavefunction remains the same after applying $m_z$ operation. 
%We will show that the $m_z$ symmetry is preserved in single-particle regime, while in the presence of Coulomb interactions, the $m_z$ symmetry could be broken spontaneously. 

%[STOP HERE] !!!!!!!!!!!!!!!
%%%%%%%%%%%%%%%%%%%%%%%%%%%%%%%%%%%%%%%%%%%%%%%%%%%%%%%%%%%
%%%%%%%%%%%%   orbital  multiferro and OME   %%%%%%%%%%%%%%%%%%%%%%%%%%%%%%%
%%%%%%%%%%%%%%%%%%%%%%%%%%%%%%%%%%%%%%%%%%%%%%%%%%%%%%%%%%%
\section{Correlation effects in mirror-symmetric ATMG system} %The bandwidth around CNP is small, which indicates a strong Coulomb interaction in this . We consider long-range Coulomb interaction in ATMG\cite{zhangshihao-sp}:
%According to the partition rules, there are bands with small bandwidth in ATMG, which indicates a strong Coulomb interaction in this system. Here we consider long-range Coulomb interaction, which is expressed as: 
%The various  low-energy band structures 
%As the ATMG systems have flat bands m
The ubiquitous flat bands in  ATMG  make these systems strongly susceptible to Coulomb interactions. Moreover, unlike magic-angle TBG, in magic-angle ATMG typically there are  flat bands co-existing with other dispersive bands (such as Dirac cone)  or  double flat bands in these systems. The extra low-energy dispersive bands (e.g., Dirac cone) may be coupled with the flat bands under weak displacement fields and display different correlated states from those in magic-angle TBG at certain filling factors \cite{lian-trilayer2-prb21,bernevig-trilayer1-prb21}.  On the other hand, in mirror-symmetric ATMG  with double flat bands, e.g., in $A$-$ABA$-$A$ system, the extra pair of flat bands marked by opposite mirror eigenvalues introduce additional degrees of freedom. What are the correlated ground states in such double-flat-band systems at different filling factors of the flat bands, how the extra degrees of freedom (mirror eigenvalues) would play a role, and how the correlated states would differ from those of TBG, are all open questions. We try to answer these questions by studying the correlated states at different integer fillings of  ATMG with $A$-$ABA$-$A$ stacking, the simplest mirror-symmetric ATMG system with double flat bands. 

\subsection{Symmetry-breaking ground states}
We have performed unrestricted self-consistent Hartree-Fock calculations for the $A$-$ABA$-$A$ system within the subspace of the double flat bands. In particular, the intersite Coulomb interactions in graphene-based systems can be written as 
\begin{equation}
\begin{split}
H_{C}=\frac{1}{2N_{s}}&\sum_{\alpha\alpha^{\prime}}\sum_{\textbf{k}_{a}\textbf{k}^{\prime}_{a}\textbf{q}_{a}}\sum_{\sigma\sigma^{\prime}}V(\textbf{q}_{a}) \\
&\times\hat{c}^{\dagger}_{\textbf{k}_{a}+\textbf{q}_{a},\alpha\sigma}\hat{c}^{\dagger}_{\textbf{k}_{a}^{\prime}-\textbf{q}_{a},\alpha^{\prime}\sigma^{\prime}}\hat{c}_{\textbf{k}_{a}^{\prime},\alpha^{\prime}\sigma^{\prime}}\hat{c}_{\textbf{k}_{a},\alpha\sigma}
\end{split}
\end{equation}
where $N_s$ is the total number of moir\'e cells in the system, $\textbf{k}_{a}, \textbf{k}^{\prime}_{a}, \textbf{q}_{a}$ are the atomic wavevectors of graphene, $\alpha$, $\alpha'$ denote the layer and sublattice indices, and $\sigma$ is the spin index. If one expands the wavevector around the Dirac points $\mathbf{K}$ and  $\mathbf{K}'$, the Coulomb interaction can be further decomposed into the intravalley part and the intervalley part, with the former interaction strength being two orders of magnitudes greater than the latter for small twist angles $\theta\!\sim\!1^{\circ}$. Therefore, we only consider the intravalley Coulomb interactions in this work, which is expressed as 
\begin{align}
H_{C}^{\textrm{intra}}=\frac{1}{2N_{s}}\sum_{\textbf{k}_{a}\textbf{k}_{a}^{\prime}\textbf{q}_{a}}\sum_{\mu\mu^{\prime},\sigma\sigma^{\prime},\alpha\alpha^{\prime}}\,&V(\textbf{q}_{a})\hat{c}^{\dagger}_{\textbf{k}_{a}+\textbf{q}_{a},\mu\alpha\sigma}\hat{c}^{\dagger}_{\textbf{k}_{a}^{\prime}-\textbf{q}_{a},\mu^{\prime}\alpha^{\prime}\sigma^{\prime}}\;\nn
&\times\hat{c}_{\textbf{k}_{a}^{\prime},\mu^{\prime}\alpha^{\prime}\sigma^{\prime}}\hat{c}_{\textbf{k}_{a},\mu\alpha\sigma}
\end{align}
where $\mu,\mu^{\prime}=\pm$ are the valley indices, and now the wavevectors $\textbf{k}$, $\textbf{k}^{\prime}$, and $\textbf{q}$ are expanded around the Dirac point of valley $\mu$ ($\mathbf{K}^{\mu}$), which can decomposed as $\textbf{k}_{a}=\textbf{k}+\textbf{Q}$, where $\textbf{k}$ is the moir\'e wavevector in moir\'e Brillouin zone, and $\textbf{Q}$ is the moir\'e lattice vector. 
A single-gate screened Coulomb interaction $V(\textbf{q}_{a})=e^{2}(1-e^{-2\lvert \textbf{q}_{a}\rvert d_{s}})/(2\Omega_{M}\epsilon_{\rm{BN}}\lvert \textbf{q}_{a}\rvert)$ is adopted in this work, where $\Omega_{M}$ is the area of moir\'e primitive cell, $d_{s}\!=\!40\,$nm  is the distance between ATMG layers and the metallic gate, $\epsilon_{\rm{BN}}\!\approx\!4$ is the  dielectric constant of the BN substrate. The Coulomb interactions are further projected onto the double flat bands,  and we perform self consistent Hartree-Fock calculations within the subspace of the double flat bands. Besides, the Coulomb interactions between electrons in  flat bands can be further screened by remote-band particle-hole excitation, and such screening effects in our calculation are treated with the constrained random phase approximation(cRPA) \cite{zhang-tbghf-arxiv20}, where the cRPA dielectric constant $\epsilon(\textbf{q}+\textbf{Q})=\epsilon_{\rm{BN}}(\mathbbm{1}+\hat{\chi}^{0}(\textbf{q})\hat{V}(\textbf{q}))_{\textbf{Q},\textbf{Q}}$, where $\hat{\chi}^{0}(\textbf{q})$ is the zero-frequency bare susceptibility at moir\'e wavevector $\textbf{q}$, and $\hat{V}(\textbf{q})$ is the Coulomb interaction matrix defined in the space of reciprocal moir\'e lattice vector $\mathbf{Q}$, with $\hat{V}(\textbf{q})_{\mathbf{Q},\mathbf{Q}}=V(\textbf{q}+\mathbf{Q})$. We refer the readers to Supplementary Information (Appendix III-IV) for more details about the Hartree-Fock and cRPA formalism.

%In this work, we consider Coulomb interaction in $A$-$ABA$-$A$, the simplest mirror-symmetric ATMG system with double flat bands. We project the Coulomb interaction onto double flat bands and perform Hartree-Fock self-consistant calculation with a cRPA calculation. 
To start with, we calculate the ground states at different integer fillings of the double flat bands using the Hartree-Fock and cRPA methods described above. The filling factor is counted with respect to the CNP, i.e., the filling factor is defined as $\nu\!=\!n-8$ when $n$ out of the 16 flat bands (including valley and spin degeneracy) are filled. Then we calculate the expectation values of the order parameters of the Hartree-Fock ground states at each integer filling, and figure out the dominant ones which are presented in Table~\ref{tab:filling}, where $\mathbf{\tau}$, $\mathbf{s}$, and $\mathbf{\sigma}$ denote Pauli matrices defined in valley, spin, and sublattice space respectively.  For example, the ground state at filling -3 is a gaped spin-valley polarized state. In order to depict the spontaneous $m_z$ symmetry breaking, we also calculate the vertical  electric polarization at different filling factors. The  vertical electric polarization per moir\'e supercell $p_z$ is defined as: $p_{z}\!=\!\sum^{5}_{l=1}\,(l-3)\,q_{l}d_{0}$, where $d_{0}\!=\!3.35\,\angstrom$ is the interlayer distance of Bernal bilayer graphene, $q_{l}=e\left\langle \tau_{0}\s_{0}\mathbb{L}_{l}\sigma_{0}\right\rangle$ is the layer resolved charge density, where $\mathbb{L}_l$ is the projection operator onto  layer $l$, a $5\times 5$ matrix with the $l$th diagonal element identity and all other elements being zeros. %defined in layer subspace that the only identity element on the l-th column, l-th row with other elements be zero. 
The unit of the electric polarization is $e\cdot\angstrom$ per moir\'e supercell. We also evaluate the orbital magnetization and valley polarization.  The valley polarization $\xi_z$ is defined as: $\xi_z\!=\!\sum_{l=1}^{5}\xi_z(l)\!=\!\sum_{l=1}^{5}\left\langle\tau_{z}s_{0}\mathbb{L}_l\sigma_{0}\right\rangle$, where $\langle\tau_z(l)\rangle$ is defined as the valley polarization projected onto layer $l$. A finite valley polarization would split the two valleys and would give rise to nonzero net orbital magnetization. In Table~\ref{tab:filling}, we present the calculated vertical electric polarization and valley polarization of the spontaneous symmetry-breaking states at different filling factors. We find that $m_z$ symmetry is spontaneously broken by Coulomb interactions at all integer fillings, which generate small but nonzero  electric polarization. %The  vertical polarization is larger when the filling factor is between -2 to +6, since at these fillings more electrons are accumulated distributes on two side instead of the middle layer. This can be understood through a single-particle effect. For the rodust flat bands, electrons distribute only on two side, the distribution of middle layer comes from the lifted flat bands. 

\begin{table}[]\scriptsize
    \centering
    \caption{Gap, main order parameter, polarization and valley polarization of the Hartree-Fock ground states at different filling factor. The unit of polarization is $e\cdot\angstrom$ per moir\'e supercell. The unit of valley polarization is charge per unit cell.}
    \begin{tabular}{c|c|c|c|c}
    \hline\hline
     filling factor   & gap(eV) & main order  & polarization &valley \\
      &  & parameter & &polarization\\
      \hline
       -7 & 0.0119 & $s_{z}\tau_{0}\sigma_{0}$  & -0.0103& 0.398 \\
       -6 & 0.0208& $s_{z}\tau_{0}\sigma_{0}$  & -0.0009 & 0.048\\
       -5 & 0.0132& $s_{z}\tau_{0}\sigma_{0}$, $s_{0}\tau_{z}\sigma_{0}$  & -0.0110 & 0.624 \\
       -4& / & $s_{z}\tau_{0}\sigma_{0}$   & 0.0197 & -1.179\\
       -3   & 0.0148& $s_{z}\tau_{0,z}\sigma_{0}$ & -0.0112 & -0.406 \\
       -2 &/& $s_{0}\tau_{z}\sigma_{0}$  & 0.0732 & -1.364 \\
       -1  & 0.0073 & $s_{0}\tau_{z}\sigma_{0}$ & 0.0585 & -3.046 \\
       0  & /& $s_{0}\tau_{x}\sigma_{y}$ & 0.1004 & 0.014 \\
       1 & / & $s_{0}\tau_{z}\sigma_{0}$  & 0.0708 & -2.705 \\
       2  &0.0095 & $s_{z}\tau_{0}\sigma_{0}$ & 0.0834 & -0.030 \\
       3 & 0.0063 & $s_{0,z}\tau_{0,z}\sigma_{0}$  & 0.0713 & 1.000 \\
       4 &0.0131 & $s_{0}\tau_{0}\sigma_{0}$ & 0.0834 & 0.000 \\
       5  & / & $s_{0,z}\tau_{0,z}\sigma_{0}$ & 0.0771 & -0.988 \\
       6 & / & $s_{0}\tau_{z}\sigma_{0}$ & 0.0741 & -1.963 \\
       7  & / & $s_{0,z}\tau_{0,z}\sigma_{0}$ & 0.0264 & -0.967 \\
       \hline\hline
    \end{tabular}
    \label{tab:filling}
\end{table}

\begin{figure}[b]
\begin{center}
    \includegraphics[width=8.5cm]{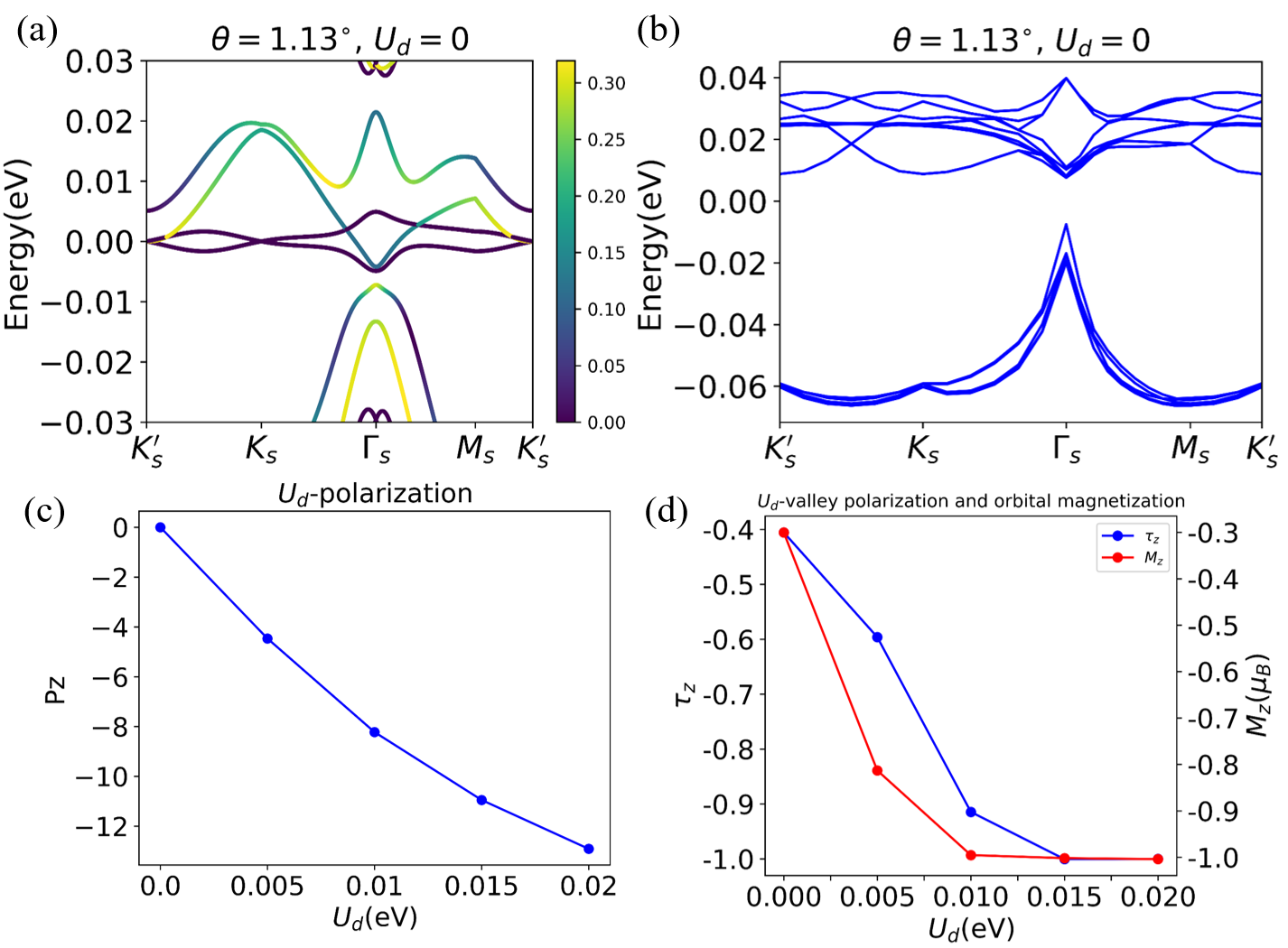}
\caption{
The band structure of $A-ABA-A$ for two valleys including intrasublattice moir\'e potential and the further neighbor interlayer hopping. (a)The non-interacting band structures with weight projected to the middlemost layer. (b) The Hartree-Fock band structure for two valleys. (c) The vertical polarization, and (d) the valley polarization and the corresponding orbital magnetization of the symmetry-breaking ground state at -3 filling as a function of vertical electrostatic potential drop $U_d$.}% The solid lines with points are the calculated results, the dashed lines are the results obtained through symmetry operations.}
\label{fig:131}
\end{center}
\end{figure}

\subsection{Orbital magnetoelectric effect through intertwined orders} 
To characterize the effect of vertical electric field, we calculate both the vertical electric polarization and valley polarization of the Hartree-Fock ground states for the $A-ABA-A$ system at filling factor -3 with increasing displacement fields. The displacement  field ($D$) is introduced by applying a homogeneous vertical electrostatic potential difference $U_d$ between the topmost and bottommost layer%two adjacent layers
, i.e., $U_d=eD d_{0}\times 4/\epsilon_{\rm{BN}}$, where $\epsilon_{\rm{BN}}\approx 4$ is the dielectric constant of the BN substrate. 
Our calculations indicate that the dominant order parameters of the ground states at filling -3 are unchanged under different $U_d$, i.e., the system always  stays in the spin-valley polarized state with broken $m_z$ symmetry, suggesting that the no phase transition occurs at least for $0\!\leq\!U_d\!\leq\!0.02\,$eV.  However, by virtue of the $m_z$ symmetry and the additional layer degrees of freedom, the valley polarization acquires nontrivial layer distributions as shown by the $\xi_z(l)$ ($l=1,...,5$) values in Table~\ref{tab:vz}.  We see that $\xi_{z}(1)$ and $\xi_z(2)$ are approximately the same, while $\xi_{z}(4)$ and $\xi_z(5)$ are approximately  the same, which are different from $\xi_z(3)$.  As a result, the layer-resolved valley polarization can be approximately decomposed into three terms:
%\begin{widetext}
\begin{align}
\hat{\xi_z} \approx \xi_z^{s} \,\mathbbm{1}_{5\times 5}+ 
                   \xi_{z}^{a} \hat{P}_z +  \xi_{z}^{a} \hat{Q}_z 
\end{align}
where 
\begin{align}
\hat{P}_{z}=\left(\begin{array}{ccccc}
				-2& 0& 0 &0&0  \\
				0&-1&0&0&0\\
				0&0&0&0&0\\
                         0&0&0&1&0\\
                          0&0&0&0&2 \end{array}\right)\;,\;\;  \hat{Q}_{z}=\left(\begin{array}{ccccc}
				1& 0& 0 &0&0  \\
				0&0&0&0&0\\
				0&0&0&0&0\\
                0&0&0&0&0\\
                0&0&0&0&-1 \end{array}\right)
\end{align}
and $\hat{\xi}_z$ is a $5\times 5$ matrix with its $l$th diagonal element denoting the valley polarization contributed by layer $l$, i.e., $\hat{\xi}_{z,ll}=\xi_{z}(l)$.
$\xi_z^{s}$ and $\xi_z^{a}$ denote the layer-symmetric and layer anti-symmetric components of the valley polarization, with $\xi_z^{s}=\sum_{l=1}^{5}\xi_z(l)/5$, and $\xi_z^{a}=\sum_{l\neq 3} \rm{sgn}[(l-3)](\xi_z(l)-\xi_z^{s})/4$. In other words, $\xi_z^{s}=\langle \tau_z\otimes \mathbbm{1}_{5\times 5}\rangle/5$ is the layer average of valley polarization ($\mathbbm{1}_{5\otimes 5}$ denotes identity matrix in layer space), and $\xi_z^{a}=\langle \tau_z\otimes (\hat{P}_z+\hat{Q}_z) \rangle /4$, where $\langle \hat{O} \rangle$ denotes the expectation value of operator $\hat{O}$ evaluated with respect to the symmetry-breaking ground state. The layer-symmetric and layer anti-symmetric valley polarization for the ground states  at filling $-3$ with different $U_d$ are presented in the last two rows of Table~\ref{tab:vz}.

We note that $\hat{P}_z$ is exactly the vertical electric polarization operator $\hat{p}_z\!=\!e d_0\hat{P}_z$, which couples linearly to external electric field; while the  valley polarization operator is proportional to the orbital magnetization operator $\hat{M}_z\!=\!g_z\mu_B\tau_z$ which couples linearly to external magnetic field, where $g_z$ is introduced as an effective $g$ factor and $\mu_B$ is the Bohr magneton. 
%It follows that one can use a simplified Hamiltonian to describe the valley polarized and $m_z$-broken state at filling $-3$ and its response external electric and magnetic fields:
%So that we can construct simplified model with homogeneous order parameter in reciprocal space and assume the state is a pure valley polarized state with a only dependence in layer subspace under finite vertical electric field:
Given the above discussions, we introduce an effective mean-field Hamiltonian to describe how the symmetry-breaking state at filling -3 would respond to external electric and magnetic fields
\begin{equation}
\begin{split}
H(\textbf{k})=&H_{0}(\textbf{k})+ \Delta_{z0}\,\tau_{z}\otimes\mathbbm{1}_{5\times5}+\Delta_{zz}\,\tau_{z}\otimes{(\hat{P}_z + \hat{Q}_z)}\\
              & -g_z\mu_B B_z\,\tau_{z}\otimes\mathbbm{1}_{5\times5} + \frac{1}{4}U_d \,\tau_{0}\otimes \hat{P}_{z}\;,
\end{split}
\label{eq:order}
\end{equation}
where $\Delta_{z0}$ and $\Delta_{zz}$ are the ``mean fields" that are self consistently generated by Coulomb interactions which are coupled with the $\tau_{z}\otimes\mathbbm{1}_{5\times5}$ operator and $\tau_{z}\otimes{(\hat{P}_z + \hat{Q}_z)}$ operator respectively, while $B_z$ is the vertical magnetic field and $U_d$ is the vertical electrostatic energy drop.
%It indicates that the valley polarization $v_z$ can be tuned by vertical electric field and the polarization $p_z$ can be tuned by vertical magnetic field.  To be specific, the layer symmetric term  of order parameter is coupled to magnetic field, while the layer anti-symmetric term is coupled to electric field. 
%Eq.~(\ref{eq:order}) indicates that a tunable polarization and valley polarization. 
As the electric polarization operator and the valley polarization operator are intertwined together,  Eq.~(\ref{eq:order}) implies a tunable electric polarization by magnetic field and conversely a tunable valley polarization (orbital magnetization) by electric field. 
To be specific, a vertical electric field is coupled to the $\hat{P}_z$ operator, which is in turn intertwined with the layer anti-symmetric component of the valley polarization operator, thus would change the valley polarization and orbital magnetization of the system. Conversely, a vertical magnetic $B_z$ is coupled to orbital magnetization (valley polarization), and the valley polarization operator is intertwined with the electric polarization operator, which would change electric polarization of the system.
%This argument remains correct for an enhanecd vertical magnetic field. 
It is clearly seen from Table~\ref{tab:vz} that the layer symmetric valley polarization is larger than the layer anti-symmetric one, implying that the orbital magnetization still has the strongest coupling to magnetic field, but can tuned by electric field. In Fig.~\ref{fig:131}(c)and (d), we present the calculated electric polarization  and valley polarization of the symmetry-breaking states at filling -3 under different $U_d$. As $U_d$ increases, clearly the electric polarization is linearly enhanced as shown in Fig.~\ref{fig:131}(c). On the other hand, the valley polarization  and the corresponding orbital magnetization  are also dramatically   enhanced with the increase of $U_d$ as shown by the blue and red dots in Fig.~\ref{fig:131}(d). 
This indicates a novel type of  orbital magnetoelectric effect driven by Coulomb interactions in mirror-symmetric ATMG system with double flat bands.

%Since the valley polarization state indicates an energy shift between two valleys, we can apply an opposite constant energy shift to two valley of single particle flat bands to evaluate the orbital magnetization and valley polarization. %The energy scale of flat bands energy shift is 1meV. Then the orbital magnetization can be tuned by vertical displacement electrostatic field. When applying vertical magnetic fields, it will induce a non-zero orbital magnetization, indicating a energy shift between two valleys. As a result, the valley polarization $v_z$ will induce a polarization $p_z$ among different layers.That is magnetoelectric effect through orbital wavefuntion. 
%We provide the valley polarization $v_z$ and the corresponding orbital magnetization in Fig.~\ref{fig:131}(d). The solid line with points is the calculated results, while we can get the dashed lines through a $\mathcal{T}$ (time reversal) operation, which flips the orbital magnetization and valley polarization as well; $m_z$ operation, which flips the vertical electric field; and a $m_z\mathcal{T}$ operation, flipping vertical electric field, orbital magnetization and valley polarization. That is the magnetoelectric effect through orbital channel.

\begin{table}[]%\scriptsize
    \centering
    \caption{Layer resolved valley polarization for the ground states at filling -3 under different $U_d$. $\xi_z^{s}=\sum_{l=1}^{5}\xi_z(l)/5$, and $\xi_z^{a}=\sum_{l\neq 3} \rm{sgn}[(l-3)](\xi_z(l)-\xi_z^{s})/4$.}
    %\caption{The layer symmetric term of valley polarization order parameter is defined as: $v^{s}_{z}=\sum_{l\neq3}v_{z}(l)/4$. The anti symmetric one is defined as: $v^{a}_{z}=\sum_{l\neq3}\mid v_{z}(l)-v^{s}_{z}\mid/4$.}
    \begin{tabular}{lccccc}
    \hline\hline
     %\diagbox{$l$}{$U_d$}  & 0 & 0.005 & 0.010 & 0.015 & 0.020\\
      $U_d$ (eV)  & 0 & 0.005 & 0.010 & 0.015 & 0.020\\
      \hline
       $\xi_z(1)$  & -0.0994 & -0.1690 &-0.2556 &-0.2724&-0.2792\\
       \hline
       $\xi_z(2)$  & -0.0996 & -0.1676 & -0.2547 &-0.2720&-0.2780\\
       \hline
       $\xi_z(3)$  & -0.0065 & 0.0023 & -0.0685 &-0.1317&-0.1469\\
       \hline
       $\xi_z(4)$ & -0.1003 & -0.1309 & -0.1683 &-0.1626&-0.1482\\
       \hline
       $\xi_z(5)$  & -0.1000 & -0.1309 & -0.1674 &-0.1617&-0.1477\\      
       \hline
       $\xi^{s}_{z}$  & -0.0998 & -0.1496 &-0.2115 &-0.2172&-0.2133\\
       \hline
       $\xi^{a}_{z}$  & 0.0003 & 0.0187 & 0.0437 &0.0550&0.0653\\
       \hline\hline
    \end{tabular}
    \label{tab:vz}
\end{table}

%To summarize, in this work we have theoretically studied the low energy band dispersion and the Coulomb interaction effects in ATMG systems.  %a class of alternating twisted multilayer graphene in both single particle and interaction regime. 
%We have obtained generic partition rules to describe the low energy band structures. We constructed the simplified $\textbf{k}\cdot \textbf{p}$ model in analytical solution approach under chiral limit and explained the partition rules with the model. %We pointed out that ATMG with $S_{L}>$2 has double flat bands. 
%We pointed out that all the ATMG with mirror symmetry $m_z$ and $S_{L}>1$ has double flat bands. We have shown that the coulomb interaction can spontaneously break $m_z$ symmerty in mirror-symmetric ATMG such as $A-ABA-A$. We have also provided the orbital magnetization as a function of vertical electric field in $A-ABA-A$ at filling factor -3. Our results point the orbital magnetization can be tuned linearly by the vertical electric field, indicating the orbital magnetoelectic effect.

To summarize, in this work we have theoretically studied the electronic structures and interaction effects of alternating twisted multilayer graphene (ATMG) systems. We find that these ATMG systems exhibit various intriguing non-interacting band dispersion including one pair of flat bands, one pair of flat bands co-existing with Dirac cones or more generally $E(\mathbf{k})\sim k^{J}$($J$ is positive integer) dispersion, as well as two pairs of flat bands which may also co-exist with $E(\mathbf{k})\sim k^{J}$($J$ is positive integer)dispersion. Based on an analytic analysis from a simplified $\textbf{k}\cdot\textbf{p}$ model approach, we find that the low energy band structure can be described by a set of generic partition rules. 
%It follows from the partition rules that there must be two pairs of flat bands (per valley per spin) for ATMG  with mirror $m_z$ symmetry and with more than one layer in the middle sequence. %We take $A-ABA-A$('$ABA$' is Bernal stacking) as a special case. 
We  have also considered Coulomb interaction effects in ATMG with $A-ABA-A$ stacking, the simplest mirror symmetric ATMG system having two pairs of flat bands. We have studied the symmetry-breaking ground states at different integer filling factors under zero external fields based on unrestricted Hartree-Fock calculations. We find that at certain fillings both time-reversal symmetry and the mirror symmetry  can be broken spontaneously by Coulomb interactions, leading to insulator states with intertwined electric polarization and orbital magnetization. As a result of such intertwined ordering, the system can exhibit a novel type of orbital magnetoelectric effect with the orbital magnetization (electric polarization) being highly tunable by external electric (magnetic) field. Our work is a significant step forward in understanding the electronic structures and correlation effects of alternating twisted graphene systems, and will provide useful guidelines for future experimental and theoretical studies.

%%%%%%%%%%%%%%%%%%%%%%%%%%%%%%%%%%%%%%%%%%%%%%%%%%%%%%%%%%%%%%%%%%
%%%%%%%%%%%   citation  %%%%%%%%%%%%%%%%%%%%%%
%%%%%%%%%%%%%%%%%%%%%%%%%%%%%%%%%%%%%%%%%%%%%%%%%%%%%%%%%%%%%%%%%%

\acknowledgements
This work is supported by the National Key R \& D program of China (grant no. 2020YFA0309601), the National Science Foundation of China (grant no. 12174257), and the start-up grant of ShanghaiTech University.

\bibliography{atmg}
%%%%%%%%%%%%%%%%%%%%%%%%%%%%%%%%%%%%%%%%%%%%%%%%%%%%%%%%%%%%%%%%%%
%%%%%%%%%%%%%%%%%%%%%%%%%%%%%%%%%%%%%%%%%%%%%%%%%%%%%%%%%%%%%%%%%%
%%%%%%%%%%%%%%%%%%%%%%%%%%%%%%%%%%%%%%%%%%%%%%%%%%%%%%%%%%%%%%%%%%
\widetext
\clearpage

\begin{center}
\textbf{\large Supplementary Information}
\end{center}

%%%%%%%%%%%%%%%%%%%%%%%%%%%%%%%%%%%%%%%%%%%%%%%%%%%%%%%%%%%%%%%%%%
%%%%%%%%%%%%%%%%%%%%%%%%%%%%%%%%%%%%%%%%%%%%%%%%%%%%%%%%%%%%%%%%%%
\vspace{12pt}
\begin{center}
\textbf{\large \I\ The lattice structure and continuum model for alternating twisted multilayer graphene}
\end{center}

The moir\'e pattern is formed by a small twist angle $\theta$ between two layers, where $\cos(\theta)=(3m^{2}+3m+1/2)/(3m^{2}+3m+1)$. The lattice vectors of the moir\'e supercell are: $\textbf{t}_{1}=(\sqrt{3}/2,1/2)\cdot L_{s}, \textbf{t}_{2}=(0,1)\cdot L_{s}$, where $L_{s}=a/(2\sin(\theta/2))$ is the moir\'e lattice constant, $a$ is the atomic lattice constant of graphene. We consider corrugation between twisted layers for L$>$1, modeled as\cite{koshino-prx18}:
\begin{equation}
d_{z}(\textbf{r})=d_{0}+2d_{1}\sum_{j=1}^{3}\cos(\textbf{b}_{j}\cdot\delta(\textbf{r}))
\end{equation}
where $\textbf{b}_{1}=(1,1/\sqrt{3})\cdot2\pi/a, \textbf{b}_{2}=(-1,1/\sqrt{3})\cdot2\pi/a, \textbf{b}_{3}=(0,2)\cdot2\pi/a$. $\delta(\textbf{r})$ is the local shift between two carbon atoms. $d_{0}=3.433\,\angstrom$, $d_{1}=0.0278\,\angstrom$. For L=1: 
\begin{equation}
\begin{split}
d_{z,twist}(\textbf{r})&=d_{0}+2d_{1}\sum_{j=1}^{3}\cos(\textbf{b}_{j}\cdot\delta(\textbf{r})) \\
d_{z,untwist}(\textbf{r})&=d_{AB}+d_{AA}/2-d_{z,twist}(\textbf{r})
\end{split}
\end{equation}
In this situation, the middle layer can be treated as ideally flat, so that it will not break $m_z$ symmetry. The corrugation difference does not matter to continuum model. 

The low-energy electronic structure of ATMG can be well described based on the Bistritzer-MacDonald continuum model\cite{BM}. The free Dirac fermions from the atomic $K$ and $K^{\prime}$ valleys contribute to the low energy states of ATMG. We can treat low energy states from two different atomic valleys as decoupled at the non-interacting level for small twist angle. Here we only consider interlayer hopping for two neighboring layers. The continuum model describing the $\mu=\mp$ valley is:
 \begin{align}
%\begin{split}
&H^{\mu}_{ATMG} 
= \left(\begin{array}{ccc}
H^{\mu}_{N} & \mathbb{U}_{\mu}^{\dagger}e^{-i\mu\Delta\mathbf{K}\cdot\mathbf{r}} & 0\\
\mathbb{U}_{\mu}e^{i\mu\Delta\mathbf{K}\cdot\mathbf{r}} & H^{\mu}_{L} & \mathbb{U}_{\mu}e^{i\mu\Delta\mathbf{K}\cdot\mathbf{r}}\\
0 &  \mathbb{U}_{\mu}^{\dagger}e^{-i\mu\Delta\mathbf{K}\cdot\mathbf{r}} & H^{\mu}_{M}
\end{array}\right)
%\end{split}
\label{eq:H-continuumatmg}
\end{align}
where $H^{\mu}_{N}$, $H^{\mu}_{L}$ and $H^{\mu}_{M}$ are the Hamiltonians of the untwisted bilayers. We only consider the nearest-neighbor interlayer coupling between the two untwisted layers.
\begin{align}
\begin{split}
H_{M}^{\mu}=\left(\begin{array}{ccccc}h^{\mu}_{0}(\mathbf{k}) & h_{\alpha} & 0 & 0 & \cdots \\
 h_{\alpha}^{\dagger} & h^{\mu}_{0}(\mathbf{k}) & h_{\alpha} & 0 & \cdots \\ 
0 & h_{\alpha}^{\dagger} & h^{\mu}_{0}(\mathbf{k}) & h_{\alpha} & \cdots \\ 
& & & \cdots & \end{array}\right)
\end{split}
\end{align}
where $h^{\mu}_{0}(\textbf{k})=-\hbar v_{F}(\textbf{k}-\textbf{K}_{M})\cdot\bm{\sigma}^{\mu}$, $\bm{\sigma}^{\mu}=(\mu\sigma_{x},\sigma_{y})$. $h_{\alpha}$ is the interlayer hopping matrix for different stacking chirality, where $\alpha,\alpha^{\prime}=+,-$, $h_{-}=h^{\dagger}_{+}$. If we only consider nearest neighbor hopping, the hopping matrix is:
\begin{align}
\begin{split}
h_{+}=\left(\begin{array}{cc}
0 & 0 \\
t_{\perp} & 0
\end{array}\right)
\end{split}
\end{align}
Taking further neighbor hopping for untwisted layers into account, the flat bands will be more dispersive. The hopping matrix is:
\begin{align}
\begin{split}
h_{+}=\left(\begin{array}{cc}
t_{2}f(\textbf{k}) & t_{2}f^{*}(k) \\
t_{\perp}-3t_{3} & t_{2}f(\textbf{k})
\end{array}\right)
\end{split}
\label{eq:fur}
\end{align}
we set $t_{\perp}$=0.48eV, $t_{2}$=0.21eV,$t_{3}$=0.05eV. The phase factor is $f(\textbf{k})=e^{-i\sqrt{3}k_{y}a/3}+e^{i(k_{x}a/2+\sqrt{6}k_{y}a/6)}+e^{i(-k_{x}a/2+\sqrt{3}k_{y}a/6)}$\cite{jpliu-tmg-prx}.

$\mathbb{U}$ is the interlayer hopping between two sequences, here we only consider hopping between two neighboring layers.
\begin{align}
\begin{split}
\mathbb{U}e^{-i\Delta\textbf{K}\cdot\textbf{r}}=\left(\begin{array}{ccc}0 & \cdots & 0 \\
 \vdots & \cdots & \vdots  \\ 
U(\textbf{r})e^{-i\Delta\textbf{K}\cdot\textbf{r}} & \cdots & 0 \end{array}\right)
\end{split}
\end{align}
where $U(\textbf{r})$ describes the hopping induced by moir\'e pattern.:
\begin{align}
\begin{split}
U(\textbf{r})=\left(\begin{array}{cc} u_{0}g(\textbf{r}) & u^{\prime}_{0}g(\textbf{r}-\textbf{r}_{AB}) \\
 u^{\prime}_{0}g(\textbf{r}+\textbf{r}_{AB}) & u_{0}g(\textbf{r}) \end{array}\right)
\end{split}
\end{align}
where $\textbf{r}_{AB}=(\sqrt{3}L_{s}/3,0)$. Because of the corrugation effect, the intrasublattice interlayer hopping trem $u_{0}$ is different from the intersublattice interalyer hopping term $u_{0}^{\prime}$\cite{koshino-prx18}. $g(\textbf{r})=\sum_{j=1}^{3}e^{i\textbf{q}_{j}\cdot\textbf{r}}$, and $\textbf{q}_{1}=(0,2/3)\cdot2\pi/L_{s}$, $\textbf{q}_{2}=(-1/\sqrt{3},-1/3)\cdot2\pi/L_{s}$ and $\textbf{q}_{1}=(1/\sqrt{3},-1/3)\cdot2\pi/L_{s}$. $\Delta\textbf{K}=\textbf{K}_{2}-\textbf{K}_{1}=(0,2/3)\cdot2\pi/L_{s}$ is the shift between Driac fermion from two twisted layers. 

\vspace{12pt}
\begin{center}
\textbf{\large \II\ The simplified $\textbf{k}\cdot\textbf{p}$ model}
\end{center}

Here we provide the expression of $\tilde{h}_{+}$ in Eq.~\ref{eq:tildeh}. Since we only consider the nearest neighbor hopping term in the chiral limit, the interlayer hopping matrix would couple the A(B) sublattice of one layer and B(A) sublattice. For coupling between Dirac fermion of top monolayer graphene and the zero modes from alternating twisted layers, the Dirac fermion from A(B) sublattice is coupled to two wave functions from B(A) sublattice. To be specific, we first consider the coupling between the Bloch state spinor in topmost layer, denoted as $(u_{t,A}(\textbf{r})e^{i\textbf{K}_1\cdot\textbf{r}}, u_{t,B}(\textbf{r})e^{i\textbf{K}_1\cdot\textbf{r}})^{T}$, and the zero-mode wave function $(\Psi_{\alpha s,\textbf{k}}(\textbf{r}), \Psi_{\gamma s,\textbf{k}}(\textbf{r}))^{T}$, where $s=A,B$ is the sublattice indices, and the superscript ``$T$" for transpose conjugation. %The zero modes spinor is transformed from the TBG-like basis function, which is the linear combination of the Bloch states spinor from the alternating twsited layers:
%\begin{equation}
%\begin{matrix}\left(\begin{array}{c}
%\Psi_{\bar{\gamma}A,\textbf{k}}(\textbf{r})\\
%\Psi_{\bar{\gamma}B,\textbf{k}}(\textbf{r})
%\end{array}\right)
%\end{matrix}=\begin{matrix}\left(\begin{array}{cc}
%U_{11}&U_{12}\\
%U_{21}&U_{22}
%\end{array}\right)
%\end{matrix}\begin{matrix}\left(\begin{array}{c}
%\bar{\psi}_{\gamma A,\textbf{k}}(\textbf{r})\\
%\bar{\psi}_{\gamma B,\textbf{k}}(\textbf{r})
%\end{array}\right)
%\end{matrix}
%\end{equation}
The renormalized hopping matrix between $u_{t,s}(\textbf{r})e^{i\textbf{K}_{1}\cdot\textbf{r}}$ and $\Psi_{ls,\textbf{k}}(\textbf{r})$, is expressed as:
\begin {equation}
\begin{split}
\tilde{h}^{t}_{+,s s^{\prime}}=&\begin{matrix}\left(\begin{array}{c}
u_{t,s}(\textbf{r})e^{i\textbf{K}_1\cdot\textbf{r}},\;0,\;0,\;0,\;0
\end{array}\right)
\end{matrix}\begin{matrix}\left(\begin{array}{ccccc}
0&0&h_{+}&h_{+}&0\\
0&0&0&0&0\\
h_{-}&0&0&0&h_{+(-)}\\
h_{-}&0&0&0&h_{+(-)}\\
0&0&h_{-(+)}&h_{-(+)}&0
\end{array}\right)
\end{matrix}\begin{matrix}\left(\begin{array}{c}
0\\
\Psi_{\alpha s^{\prime},\textbf{k}}(\textbf{r})\\
0\\
\Psi_{\gamma s^{\prime},\textbf{k}}(\textbf{r})\\
0
\end{array}\right)
\end{matrix}\\
=&\int d\textbf{r}\  u^{*}_{t,s}(\textbf{r})e^{-i\textbf{K}_1\cdot\textbf{r}}h_{+}\Psi_{\gamma s^{\prime},\textbf{k}}(\textbf{r})\\
=&\int d\textbf{r}\  u^{*}_{t,s}(\textbf{r})e^{-i\textbf{K}_1\cdot\textbf{r}}h_{+}(\frac{1}{\sqrt{2}}\Psi_{N s^{\prime},\textbf{k}}(\textbf{r})+\frac{1}{\sqrt{2}}\Psi_{M s^{\prime},\textbf{k}}(\textbf{r}))\\
=&\frac{1}{\sqrt{2}}\int d\textbf{r}\  u^{*}_{t,s}(\textbf{r})e^{-i\textbf{K}_1\cdot\textbf{r}}t_{\perp}\Psi_{N s^{\prime},\textbf{k}}(\textbf{r})\delta_{s,B}\delta_{s^{\prime},A}\\
=&\frac{1}{\sqrt{2}}\tilde{t}_{\perp}\delta_{s,B}\delta_{s^{\prime},A}\\
%&\ \ \ \ \ \ \ \ \ \ \ +u^{*}_{t,B}(\textbf{r})e^{-i\textbf{K}_1\cdot\textbf{r}}t_{+}U_{12}(\frac{1}{\sqrt{2}}u_{N,B}(\textbf{r})e^{i\textbf{K}_1\cdot\textbf{r}}+\frac{1}{\sqrt{2}}u_{M,B}(\textbf{r})e^{i\textbf{K}_1\cdot\textbf{r}}) \\
%=&\frac{1}{\sqrt{2}}\int d\textbf{r}u^{*}_{t,A}(\textbf{r})t_{\perp}U_{12}u_{N,B}(\textbf{r})e^{i\textbf{K}_1\cdot\textbf{r}},
\end{split}
\label{eq:hopping}
\end {equation}
where $t_{\perp}$ is the interlayer intersublattice hopping parameter between two Dirac fermions. We decompose the zero modes wave function into layer basis. Since we only consider the interlayer hopping between adjacent layers, there is zero overlap between states in topmost layer and zero modes solution in M sequence. The coupling between the zero modes wave function and states from the bottom-most layer can be evaluated in a similar way:
\begin{equation}
\begin{split}
\tilde{h}^{b}_{+,s s^{\prime}}=&\begin{matrix}\left(\begin{array}{c}
0\;,\Psi_{\alpha s,\textbf{k}}(\textbf{r}),\;0\;,\Psi_{\gamma s,\textbf{k}}(\textbf{r})\;,0\;,
\end{array}\right)
\end{matrix}
\begin{matrix}\left(\begin{array}{ccccc}
0&0&h_{+}&h_{+}&0\\
0&0&0&0&0\\
h_{-}&0&0&0&h_{+}\\
h_{-}&0&0&0&h_{+}\\
0&0&h_{-}&h_{-}&0
\end{array}\right)
\end{matrix}\begin{matrix}\left(\begin{array}{c}
0\\
0\\
0\\
0\\
u_{b,s^{\prime}}(\textbf{r})e^{i\textbf{K}_1\cdot\textbf{r}}
\end{array}\right)\end{matrix}\\
%=&\int d\textbf{r}\  u^{*}_{t,s}(\textbf{r})e^{-i\textbf{K}_1\cdot\textbf{r}}h_{+}\Psi_{\gamma s^{\prime},\textbf{k}}(\textbf{r})\\
%=&\int d\textbf{r}\  u^{*}_{t,s}(\textbf{r})e^{-i\textbf{K}_1\cdot\textbf{r}}h_{+}(\frac{1}{\sqrt{2}}\Psi_{N s^{\prime},\textbf{k}}(\textbf{r})+\frac{1}{\sqrt{2}}\Psi_{M s^{\prime},\textbf{k}}(\textbf{r}))\\
=&\frac{1}{\sqrt{2}}\int d\textbf{r}\  \Psi^{*}_{N s,\textbf{k}}(\textbf{r})\delta_{s,B}\delta_{s^{\prime},A}t_{\perp}u_{b,s^{\prime}}(\textbf{r})e^{i\textbf{K}_1\cdot\textbf{r}}\\
=&\frac{1}{\sqrt{2}}\tilde{t}_{\perp}\delta_{s,B}\delta_{s^{\prime},A}\\
%&\ \ \ \ \ \ \ \ \ \ \ +u^{*}_{t,B}(\textbf{r})e^{-i\textbf{K}_1\cdot\textbf{r}}t_{+}U_{12}(\frac{1}{\sqrt{2}}u_{N,B}(\textbf{r})e^{i\textbf{K}_1\cdot\textbf{r}}+\frac{1}{\sqrt{2}}u_{M,B}(\textbf{r})e^{i\textbf{K}_1\cdot\textbf{r}}) \\
%=&\frac{1}{\sqrt{2}}\int d\textbf{r}u^{*}_{t,A}(\textbf{r})t_{\perp}U_{12}u_{N,B}(\textbf{r})e^{i\textbf{K}_1\cdot\textbf{r}},
\end{split}
\label{eq:hopping}
\end {equation}
 The explicit expressions of zero modes spinor from two sublattice are\cite{ashvin-analytical-tbg}:
\begin{equation}
\begin{split}
\Psi_{l A,\textbf{k}}(\textbf{r})=\frac{\vartheta_{(\textbf{ka}_{1}/2\pi)-\frac{1}{6},\frac{1}{6}-(\textbf{ka}_{2}/2\pi)}((x+iy)/(a_{2,x}+ia_{2,y})\mid e^{i2\pi/3})}{\vartheta_{-\frac{1}{6},\frac{1}{6}}((x+iy)/(a_{1,x}+ia_{1,y})\mid e^{i2\pi/3})}\begin{matrix}\left(\begin{array}{c}
\bar{\psi}_{\alpha A\textbf{K}}(\textbf{r})\\
\bar{\psi}_{\gamma A\textbf{K}}(\textbf{r})
\end{array}\right),
\end{matrix}\\
\Psi_{l B,\textbf{k}}(\textbf{r})=\frac{\vartheta^{*}_{(\textbf{ka}_{1}/2\pi)-\frac{1}{6},\frac{1}{6}-(\textbf{ka}_{2}/2\pi)}((-x-iy)/(a_{2,x}+ia_{2,y})\mid e^{i2\pi/3})}{\vartheta^{*}_{-\frac{1}{6},\frac{1}{6}}((-x-iy)/(a_{1,x}+ia_{1,y})\mid e^{i2\pi/3})}\begin{matrix}\left(\begin{array}{c}
\bar{\psi}_{\alpha B\textbf{K}}(\textbf{r})\\
\bar{\psi}_{\gamma B\textbf{K}}(\textbf{r})
\end{array}\right),
\end{matrix}
\end{split}
\end{equation}
where $l=\alpha,\gamma$, $\vartheta_{a,b}(x+iy\mid\tau)=\sum_{n=-infty}^{\infty}e^{i\pi\tau(n+a)^{2}}e^{2\pi i(n+a)(n+b)}$, $\textbf{a}_{1}, \textbf{a}_{2}$ are the moir\'e supercell lattice vector.

\vspace{12pt}
\begin{center}
\textbf{\large \III\ The Coulomb interaction in the twisted graphene}
\end{center}

We consider the density-density Coulomb interaction in real space in graphene system:
\begin{equation}
H_C=\frac{1}{2}\sum _{i  j }\sum _{\alpha \beta }\sum _{\sigma \sigma '} \hat{c}^{\dagger}_{i\alpha \sigma}\hat{c}^{\dagger}_{j\beta \sigma '}U^{\alpha \beta} _{ij}\hat{c}_{j\beta\sigma '}\hat{c}_{i\alpha \sigma}\;,
\end{equation}
where $i, j$ are the atomic lattice vector indices, $\alpha, \beta$ are the sublattice and layer indices, $\sigma$ is spin index.
%
%\begin{align}
%U^{\alpha \beta \sigma , \alpha ' \beta ' \sigma '} _{ij,i'j'}
%=&\int d \mathbf{r}\,d\mathbf{r}^{\prime}\,\frac{e^2}{4\pi \epsilon \vert \mathbf{r}-\mathbf{r}^{\prime}\vert}\,\phi ^*_\alpha (\mathbf{r}-\mathbf{R}_i-\tau _\alpha)\,\phi _\beta (\mathbf{r}-\mathbf{R}_j-\tau _\beta)\;\nn
%&\times\phi ^*_{\alpha  '}(\mathbf{r}-\mathbf{R}_i'-\tau _{\alpha '})\phi _{\beta  '}(\mathbf{r}-\mathbf{R}_j'-\tau _{\beta '})\chi ^{\dagger}_\sigma \chi ^{\dagger}_{\sigma '}\chi _{\sigma '}\chi _{\sigma}\;.
%\end{align}
%
$U^{\alpha \beta} _{ij}$ is the density-density interaction between two electrons, one at site $i$ layer/sublattice $\alpha$ and the other at site $j$ layer/sublattice $\beta$. We can separate the Coulomb interaction into inter-site and on-site term. Since the charge density is quite low in moir\'e super cell for electrons from flat bands, we can neglect on-site Hubbard interaction, which is at least an order of magnitude smaller than the inter-site ones after projecting on the low-energy states on the moir\'e length scale \cite{zhangshihao-sp}. We take Fourier transformation to the real space electron field operator:
\begin{equation}
\hat{c}_{i\alpha\sigma}=\frac{1}{\sqrt{N_s}}\sum_{\textbf{k}_{a}}e^{i\textbf{k}\cdot\textbf{R}_{i}}\hat{c}_{\textbf{k}_{a}\alpha\sigma}
\end{equation}
$\textbf{k}_{a}$ is the wave vector in atomic Brillouin zone, $N_s$ is the number of atomic unit cells. We can expand the low energy states for twisted graphene around the atomic wave vector, i.e $\textbf{k}_{a}=\textbf{k}+\textbf{K}$, where $\textbf{k}$ is the wave vector in moir\'e Brillouin zone, $\textbf{K}$ is the moir\'e reciprocal lattice vector. Then the inter-site Coulomb interaction can be divided into inter-valley term and intravalley term\cite{Ashvin-nc19}:
\begin{align}
H_{C}^{\rm{intra}}=\frac{1}{2N_s}\sum_{\alpha\alpha '}\sum_{\mu\mu ',\sigma\sigma '}\sum_{\mathbf{k}_{a}\mathbf{k}_{a} '\mathbf{q}_{a}}\,V(\mathbf{q}_{a})\,
\hat{c}^{\dagger}_{\mathbf{k}_{a}+\mathbf{q}_{a},\mu \sigma \alpha} \hat{c}^{\dagger}_{\mathbf{k}_{a}'-\mathbf{q}_{a},\mu '\sigma '\alpha '}\,
\hat{c}_{\mathbf{k}'_{a},\mu '\sigma '\alpha '}\hat{c}_{\mathbf{k}_{a},\mu \sigma \alpha}\;,\nn
H_{C}^{\rm{inter}}=\frac{1}{2N_s}\sum_{\alpha\alpha '}\sum_{\mu ,\sigma\sigma '}\sum_{\mathbf{k}_{a}\mathbf{k} '_{a}\mathbf{q}_{a}}\,V(\vert\mathbf{K}-\mathbf{K}'\vert)\,
\hat{c}^{\dagger}_{\mathbf{k}_{a}+\mathbf{q}_{a},\mu \sigma \alpha} \hat{c}^{\dagger}_{\mathbf{k}'_{a}-\mathbf{q}_{a},-\mu \sigma '\alpha '}
\hat{c}_{\mathbf{k}'_{a},\mu \sigma '\alpha '}\hat{c}_{\mathbf{k}_{a},-\mu \sigma \alpha}\;.
\label{eq:h-inter}
\end{align}
To reach a realistic situation, we consider the screening effect from the device: the single-gate screened Coulomb interaction is $V(\textbf{q}_{a})=e^{2}(1-e^{-2\lvert \textbf{q}_{a}\rvert d_{s}})/(2\Omega_{M}\epsilon_{\rm{BN}}\lvert \textbf{q}_{a}\rvert)$, where $\Omega_M$ is the area of moir\'e unit cell, $d_s=400\,\angstrom$ is the distance between graphene and gate and $\epsilon_{\rm{BN}}$ is the dielectric constant of h-BN. We can evaluate the energy scale intervallley and intravalley Coulomb interaction: the typical intravalley interaction energy $V_{M}\!\approx\!25\,$meV for $\theta\approx1.2^{\circ}$, while the intervalley interaction energy $V(\mid\textbf{K}-\textbf{K}^{\prime}\mid)\sim0.35\,$meV for $\theta\approx1.2^{\circ}$. As a result, we only consider intravalley Coulomb interaction in our calculation.

We can project the interaction from the original basis to the band basis through the following transformation:
\begin{equation}
\hat{c}_{\textbf{k}_{a},\mu\alpha\sigma}=\sum_{n}C_{\mu\alpha\textbf{G},n}(\textbf{k})\hat{c}_{\mu\sigma,n\textbf{k}}
\end{equation}
where $C_{\mu\alpha\textbf{G},n}(\textbf{k})$ is the expansion coefficient in the $n$-th Bloch eigenstate at moir\'e wave vector near valley $\mu$, and the summation over the band index $n$ is restricted to the flat-band subspace. We can rewrite the intravalley interaction by applying above transformation:
\begin{equation}
H^{\rm{intra}}
=\frac{1}{2N_s}\sum _{\mathbf{k} \mathbf{k}'\mathbf{q}}\sum_{\substack{\mu\mu' \\ \sigma\sigma'}}\sum_{\substack{nm\\ n'm'}}\left(\sum _{\mathbf{Q}}\,V(\mathbf{Q}+\mathbf{q})\,\Omega^{\mu \sigma,\mu'\sigma'}_{nm,n'm'}(\mathbf{k},\mathbf{k}',\mathbf{q},\mathbf{Q})\right)\,\hat{c}^{\dagger}_{\mu\sigma,n\mathbf{k}+\mathbf{q}} \hat{c}^{\dagger}_{\mu'\sigma',n'\mathbf{k}'-\mathbf{q}}\,\hat{c}_{\mu'\sigma',m'\mathbf{k}'}\,\hat{c}_{\mu\sigma,m\mathbf{k}}
\label{eq:Hintra-band}
\end{equation}
where $\Omega ^{\mu \sigma,\mu'\sigma'}_{nm,n'm'}$ is:
\begin{equation}
\Omega ^{\mu \sigma,\mu'\sigma'}_{nm,n'm'}(\mathbf{k},\mathbf{k}',\mathbf{q},\mathbf{Q})\,
=\sum _{\alpha\alpha'\mathbf{G}\mathbf{G}'}C^*_{\mu\sigma\alpha\mathbf{G}+\mathbf{Q},n}(\mathbf{k}+\mathbf{q})C^*_{\mu'\sigma'\alpha'\mathbf{G}'-\mathbf{Q},n'}(\mathbf{k}'-\mathbf{q})C_{\mu'\sigma'\alpha'\mathbf{G}',m'}(\mathbf{k}')C_{\mu\sigma\alpha\mathbf{G},m}(\mathbf{k})
\end{equation}
We can make Hartree-Fock approximation to the intersite intravalley Coulomb interaction so that the two-particle interaction can be solved in a mean-field single particle Hamiltonian. The Hartree term is:
\begin{equation}
\begin{split}
H_H^{\rm{intra}}=&\frac{1}{2N_s}\sum _{\mathbf{k} \mathbf{k}'}\sum _{\substack{\mu\mu'\\ \sigma\sigma'}}\sum_{\substack{nm\\ n'm'}}\left(\sum _{\mathbf{Q}} V(\mathbf{Q})\Omega ^{\mu \sigma,\mu'\sigma'}_{nm,n'm'}(\mathbf{k},\mathbf{k}',0,\mathbf{Q})\right)\\
&\times \left(\langle \hat{c}^{\dagger}_{\mu\sigma,n\mathbf{k}}\hat{c}_{\mu\sigma,m\mathbf{k}}\rangle \hat{c}^{\dagger}_{\mu'\sigma',n'\mathbf{k}'}\hat{c}_{\mu'\sigma',m'\mathbf{k}'} + \langle \hat{c}^{\dagger}_{\mu'\sigma',n'\mathbf{k}'}\hat{c}_{\mu'\sigma',m'\mathbf{k}'}\rangle \hat{c}^{\dagger}_{\mu\sigma,n\mathbf{k}}\hat{c}_{\mu\sigma,m\mathbf{k}}\right)
\end{split}
\end{equation}
and the Fock term is:
\begin{equation}
\begin{split}
H_F^{\rm{intra}}=&-\frac{1}{2N_s}\sum _{\mathbf{k} \mathbf{k}'}\sum _{\substack{\mu\mu'\\ \sigma\sigma'}}\sum_{\substack{nm\\ n'm'}}\left(\sum _{\mathbf{Q}} V(\mathbf{k}’-\mathbf{k}+\mathbf{Q})\Omega ^{\mu \sigma,\mu'\sigma'}_{nm,n'm'}(\mathbf{k},\mathbf{k}',\mathbf{k}’-\mathbf{k},\mathbf{Q})\right)\\
&\times \left(\langle \hat{c}^{\dagger}_{\mu\sigma,n\mathbf{k}'}\hat{c}_{\mu'\sigma',m'\mathbf{k}'}\rangle \hat{c}^{\dagger}_{\mu'\sigma',n'\mathbf{k}}\hat{c}_{\mu\sigma,m\mathbf{k}} + \langle \hat{c}^{\dagger}_{\mu'\sigma',n'\mathbf{k}}\hat{c}_{\mu\sigma,m\mathbf{k}}\rangle \hat{c}^{\dagger}_{\mu\sigma,n\mathbf{k}'}\hat{c}_{\mu'\sigma',m'\mathbf{k}'}\right)\;.
\end{split}
\end{equation}

\vspace{12pt}
\begin{center}
\textbf{\large \IV\ Constraint random phase approximation}
\end{center}

The Coulomb interaction in previous section is screened by device, as a result, the Coulomb interaction in reciprocal space is not a Thomas-Fermi form but a single-gate screened from. In this section, we consider further screening effects in band basis. To be specific, the interaction between electrons in flat bands can be screened by virtual excitation of particle-hole pairs from the remote bands. Such screening effects are evaluated by constrained random phase approximation(cRPA).
%First, we have bare susceptibility: 
%\begin{equation}
%\chi^{0}_{\tilde{k},\mu,m,n}(\tilde{q},\nu)=\frac{f(E_{\mu,m,\tilde{k}+\tilde{q}})-f(E_{\mu,n,\tilde{k}})}{E_{\mu,n,%\tilde{k}}+\nu-E_{\mu,m,\tilde{k}+\tilde{q}}}
%\end{equation}
%where $\mu$ is vally, $\tilde{k}$ is wave vector in moire Brillouin zone, m and n are band indices. Here we only consider static susceptibility, so that $\nu=0$. Condsider bubble diagram for Coulomb interaction:
We consider the bubble diagram for screened Coulomb interaction: 
\begin{align}
V^{cRPA}_{\textbf{k}\mu n m,\textbf{k}^{\prime}\mu^{\prime} m^{\prime} n^{\prime}}(\textbf{q})&= V^{0}_{\textbf{k}\mu n m,\textbf{k}^{\prime}\mu^{\prime} m^{\prime} n^{\prime}}(\textbf{q})\\
&+ \frac{-2}{N_{k}}\sum_{n_{1}m_{1}}^{\prime}\sum_{\textbf{k}_{1}\mu_{1}}{V^{0}_{k\mu n m,\textbf{k}_{1}\mu_{1} m_{1} n_{1}}\chi^{0}_{\textbf{k}_{1}\mu_{1}m_{1}n_{1}}(\textbf{q})V^{0}_{\textbf{k}_{1}\mu_{1} m_{1} n_{1},\textbf{k}^{\prime}\mu^{\prime} n^{\prime} m^{\prime}}}\nonumber\\
&+ \left(\frac{-2}{N_s}\right)^{2}\sum_{\substack{n_{1}m_{1}\\ n_{2}m_{2}}}^{\prime}\sum_{\mu_{1}\mu_{2}}\sum_{\textbf{k}_{1}\textbf{k}_{2}}V^{0}_{\textbf{k}\mu n m,\textbf{k}_{1}\mu_{1} m_{1} n_{1}}\chi^{0}_{\textbf{k}_{1}\mu_{1}m_{1}n_{1}}(\textbf{q})V^{0}_{\textbf{k}_{1}\mu_{1} m_{1} n_{1},\textbf{k}_{2}\mu_{2} m_{2} n_{2}}\chi^{0}_{\textbf{k}_{2}\mu_{2}m_{2}n_{2}}(\textbf{q})V^{0}_{\textbf{k}_{2}\mu_{2} m_{2} n_{2},\textbf{k}^{\prime}\mu^{\prime} n^{\prime} m^{\prime}} \nonumber \\
&+ \dots \nonumber
\label{eq:crpa}
\end{align}
We can define the single-gate screened Coulomb interaction projected to flat bands subspace:
\begin{align}
V^{0}_{\textbf{k}\mu n m,\textbf{k}^{\prime}\mu^{\prime}m^{\prime}n^{\prime}}(\textbf{q})&=\sum_{\textbf{Q}}V(\textbf{q}+\textbf{Q})\lambda_{\textbf{k}\mu n m}(\textbf{q},\textbf{Q})\lambda^{*}_{\textbf{k}^{\prime}\mu^{\prime}m^{\prime}n^{\prime}}\\
\lambda_{\textbf{k}\mu n m}(\textbf{q},\textbf{Q})&=\sum_{\alpha \textbf{G}}C^{*}_{\mu \alpha \textbf{G}+\textbf{Q},n}(\textbf{k}+\textbf{q})C_{\mu \alpha \textbf{G},m} (\textbf{k}+\textbf{q})
\end{align}
where $\textbf{G}$ is the moire reciprocal lattice vector. % We define $V^{0}(\tilde{q})_{Q,Q}=U(\tilde{q}+Q)$. Then the interaction becomes:
The summation of band indices in Eq.~(\ref{eq:crpa}) is restricted: $m_1$ and $n_1$ can not both come from the flat bands subspace. The remote bands below the flat bands are filled, while remote bands above the flat bands are empty. That is to say, the virtual excitation can happen through three channel: from the remote bands below the CNP to flat bands; from flat bands to the remote bands above CNP; from remote bands below CNP to flat bands above CNP. 
%\begin{align}
%V_{\tilde{k}\mu n m,\tilde{k^{\prime}}\mu^{\prime} m^{\prime} n^{\prime}}(\tilde{q})= \sum_{Q,Q^{\prime}}\lambda_{\tilde{k}\mu n m,Q}&(\tilde{q})[ V^{0}(\tilde{q})_{Q,Q}\delta_{Q,Q^{\prime}} \\
%&+ V^{0}(\tilde{q})_{QQ}(-\chi^{0}_{Q,Q^{\prime}})V^{0}(\tilde{q})_{Q^{\prime}Q^{\prime}} \nonumber \\
%&+ \sum_{Q^{\prime\prime}}V^{0}(\tilde{q})_{Q,Q}(-\chi^{0}_{Q,Q^{\prime\prime}})V^{0}(\tilde{q})_{Q^{\prime\prime},Q^{\prime\prime}}(-\chi^{0}_{Q^{\prime\prime},Q^{\prime}})V^{0}(\tilde{q})_{Q^{\prime},Q^{\prime}}  \nonumber \\
%&+ \dots]\lambda^{\dagger}_{Q^{\prime},\tilde{k^{\prime}}\mu^{\prime}m^{\prime}n^{\prime}}(\tilde{q}) \nonumber
%\end{align}
We can define the bare susceptibility in the transferred reciprocal vectors basis $\chi^{0}_{Q,Q^{\prime}}$ as:
\begin{equation}
\chi^{0}_{\textbf{Q},\textbf{Q}^{\prime}}=\frac{2}{N_{\textbf{k}}}\sum_{\textbf{k}_{1}}\sum _{\mu_{1}m_{1}n_{1}}^{\prime}\lambda^{\dagger}(\textbf{q},\textbf{Q})_{\textbf{k}_{1}\mu_{1}m_{1}n_{1}}\chi^{0}_{\textbf{k}_{1}\mu_{1}m_{1}n_{1}}(\textbf{q})\lambda(\textbf{q},\textbf{Q}^{\prime})_{\textbf{k}_{1}\mu_{1}m_{1}n_{1}}
\end{equation}
where: 
\begin{equation}
\chi^{0}_{\textbf{k},\mu,m,n}(\textbf{q},\nu)=\frac{f(E_{\mu,m,\textbf{k}+\textbf{q}})-f(E_{\mu,n,\textbf{k}})}{E_{\mu,n,\textbf{k}}+\nu-E_{\mu,m,\textbf{k}+\textbf{q}}}
\end{equation}
where $\mu$ is valley, $\textbf{k}$ is wave vector in moir\'e Brillouin zone, m and n are band indices. Here we only consider static susceptibility, i.e $\nu=0$.
%Note that the sum over band indices $m_{1},n_{1}$ is restricted: $m_{1},n_{1}$ can not both from the flat bands.That's the reason why we call it constrained RPA.
%Also we can define effective interaction:
%\begin{align}
%V^{eff}(\tilde{q})= \sum_{Q,Q^{\prime}}\lambda^{\dagger}_{\tilde{k}\mu n m,Q}&(\tilde{q})V_{\tilde{k}\mu n m,\tilde{k^{\prime}}\mu^{\prime} m^{\prime} n^{\prime}}(\tilde{q})\lambda_{Q^{\prime},\tilde{k^{\prime}}\mu^{\prime}m^{\prime}n^{\prime}}(\tilde{q})
%\end{align}
We can rewrite the screened Coulomb interaction in matrix form:
\begin{align}
\hat{V}^{cRPA}(\textbf{q})&=\hat{V}(\textbf{q})_{\textbf{Q},\textbf{Q}}\delta_{\textbf{Q},\textbf{Q}^{\prime}} \\
&+ \hat{V}(\textbf{q})_{\textbf{Q}\textbf{Q}}(-\chi^{0}_{\textbf{Q},\textbf{Q}^{\prime}})\hat{V}(\textbf{q})_{\textbf{Q}^{\prime}\textbf{Q}^{\prime}} \nonumber \\
&+ \sum_{\textbf{Q}^{\prime\prime}}\hat{V}(\textbf{q})_{\textbf{Q},\textbf{Q}}(-\chi^{0}_{\textbf{Q},\textbf{Q}^{\prime\prime}})\hat{V}(\textbf{q})_{\textbf{Q}^{\prime\prime},\textbf{Q}^{\prime\prime}}(-\chi^{0}_{\textbf{Q}^{\prime\prime},\textbf{Q}^{\prime}})\hat{V}(\textbf{q})_{\textbf{Q}^{\prime},\textbf{Q}^{\prime}}  \nonumber \\
&+ \dots \nonumber \\
&=\hat{V}^{0}(\textbf{q})\cdot(\mathbbm{1}+\chi^{0}(\textbf{q})\cdot \hat{V}^{0}(\textbf{q}))^{-1}
\end{align}
We define the dielectric matrix as:
\begin{align}
\hat{\epsilon}^{cRPA}(\textbf{q})_{\textbf{Q},\textbf{Q}^{\prime}}=(\mathbbm{1}+\chi^{0}(\textbf{q})\cdot V^{0}(\textbf{q}))_{\textbf{Q},\textbf{Q}^{\prime}}
\end{align}
We can make Hartree-Fock approximation to the screened Coulomb interaction and decomposed into two terms. For the Hartree term, we take $\hat{V}^{cRPA}(\textbf{q}=0)$, while we take $V(\textbf{k}^{\prime}-\textbf{k}+\textbf{Q})/\epsilon(\textbf{k}^{\prime}-\textbf{k}+\textbf{Q})$ for the Fock term.

\end{document}